\documentclass[twocolumn,prb,showpacs,multicol,amsmath,amssymb]{revtex4-1}
\usepackage{amssymb}
\usepackage{amsmath}
\usepackage{graphicx}
\usepackage{epstopdf}
\usepackage{bm}%
\usepackage{gensymb}
\usepackage{soul}
\usepackage{sidecap}
\usepackage[normalem]{ulem}

\newcommand{\bra}[1]{\ensuremath{\langle #1 |}}
\newcommand{\ket}[1]{\ensuremath{| #1 \rangle}}

\newcommand{\be}{\begin{equation}}
\newcommand{\ee}{\end{equation}}
\newcommand{\bk}{{{\bf{k}}}}

\newcommand{\bg}{{{\bf{g}}}}
\newcommand{\bH}{{{\bf{H}}}}
\newcommand{\bM}{{{\bf{M}}}}

\newcommand{\hh}{{\hat{h}}}

\newcommand{\bea}{\begin{eqnarray}}
\newcommand{\eea}{\end{eqnarray}}

\newcommand{\bd}{\begin{displaymath}}
\newcommand{\ed}{\end{displaymath}}
\newcommand{\ba}{\begin{array}}
\newcommand{\ea}{\end{array}}
\newcommand{\bi}{\begin{itemize}}
\newcommand{\ei}{\end{itemize}}
\newcommand{\bc}{\begin{center}}
\newcommand{\ec}{\end{center}}
\newcommand{\bfl}{\begin{flushleft}}
\newcommand{\efl}{\end{flushleft}}
\newcommand{\bfr}{\begin{flushright}}
\newcommand{\efr}{\end{flushright}}
\newcommand{\no}{\nonumber}

\newcommand{\mi}{\rm i}
\newcommand{\DsH}{Dresselhaus}

\def\ket#1{\left\vert #1 \right\rangle}

\def\bk{{\bf k}}   
\def\bg{{\bf g}}  \def\bd{{\bf d}}

\def\6{\partial}

\def\bra{\langle}
\def\ket{\rangle}
\def\={\!\!\!&=&\!\!\!}
\def\+{\!\!\!&&\!\!\!+~}
\def\-{\!\!\!&&\!\!\!-~}

 \newcommand\redout{\bgroup\markoverwith{\textcolor{red}{\rule[.5ex]{2pt}{0.4pt}}}\ULon}

\usepackage[colorlinks=true,citecolor=blue]{hyperref}
\hypersetup{colorlinks=true,citecolor=blue,linkcolor=red,urlcolor=blue}

 \graphicspath{{.}{./Figs/}}

\usepackage{hyperref,cleveref}
\begin{document}
\title{
 Edge currents  as a  probe of the strongly spin-polarized topological noncentrosymmetric superconductors}

\author{Mehdi Biderang$^{1,2}$} %
\author{Heshmatollah Yavari$^{1}$}\email{h.yavary@sci.ui.ac.ir}
\author{Mohammad-Hossein Zare$^{3}$}
\author{Peter Thalmeier$^{4}$}
\author{Alireza Akbari$^{2,5,6}$} \email{alireza@apctp.org}
\affiliation{$^1$Department of Physics, University of Isfahan, Hezar Jerib, 81746-73441, Isfahan, Iran}
\affiliation{$^2$Asia Pacific Center for Theoretical Physics, Pohang, Gyeongbuk 790-784, Korea}
\affiliation{$^3$Department of Physics, Faculty of Science, Qom University of Technology, Qom 37181-46645, Iran}
\affiliation{$^4$Max Planck Institute for Chemical Physics of Solids, D-01187 Dresden, Germany}
\affiliation{$^5$Department of Physics, POSTECH, Pohang, Gyeongbuk 790-784, Korea}
\affiliation{$^6$Max Planck POSTECH Center for Complex Phase Materials, POSTECH, Pohang 790-784, Korea}
\date{\today}
%
\begin{abstract}
Recently the  influence  of antisymmetric spin-orbit coupling has been studied 
in novel topological  superconductors such as half-Heuslers and artificial hetero-structures.
We investigate the effect of  Rashba and/or \DsH~spin-orbit couplings on the  band structure and topological properties of a two-dimensional noncentrosymetric   superconductor.
For this goal, the topological helical edge modes are analyzed for different spin-orbit couplings as well as for  several superconducting pairing symmetries.
To explore  the 
 transport properties, we examine the response of the spin-polarized edge states to an exchange field in a superconductor-ferromagnet  heterostructure.
 The broken chiral symmetry causes
   the
     uni-directional currents at opposite edges.
\end{abstract}
%
\maketitle
%
\section{Introduction}
Discovery of  novel 
 phases in the nodal systems has extended the classification of  states of matter from the bulk gapped 
topological insulators to the gapless systems such as topological superconductors~\cite{Zahid:2010,Qi:2011,Sato:2017,Zare:2017,Lambert:2017} 
and semimetals~\cite{Wan:2011,Turner:2013,Yan:2017,Zahid:2017,Biderang:2018aa}.
The former are new quantum states 
 with unconventional pairing symmetries exhibiting edge modes.
These zero-energy gapless modes can host Majorana fermions, which obey non-Abelian statistics unlike bosons and fermions.
These exotic edge modes are topologically protected against disorders and perturbations that gives them many promising applications~\cite{Nayak:2008,Terhal:2015,Alicea:2010}. \\

One of the most prominent platforms to realize the topological superconductivity is the class of noncentrosymmetric superconductors
(NCSs)~\cite{Bauer:2012,Schnyder:2015,Sato:2006,Qi:2010,Sato:2011,Schnyder:2011,Yip:2014,Yada:2011,Schnyder:2012,Dahlhaus:2012,Matsuura:2013,Daido:2016aa,Daido:2017aa}.
These materials lack inversion symmetry and are characterized by strong antisymmetric spin-orbit coupling (SOC), which induces a non-trivial topology for the electronic band structure~\cite{Timm:2015}. 
 This leads to the existence of helical Majorana modes, zero-energy flat-bands~\cite{Tanaka:2012,Matsuura:2013,Schnyder:2015}  and arc surface states~\cite{Tang:2014,Kai:2014}. 
 Recently attention has been attracted by the influence of SOC  on  structure of superconducting gap function and topological nature of superconductors, particularly  in electronic and spintronic device 
 design~\cite{Beenakker:2013,Beenakker:2015,Sato_Fujimoto:2009,Sato:2010,Wong:2013,Read:2000,Law:2009,Alicea:2010}.
The topological nature of NCSs and consequently the properties of the accompanied surface states can be controlled by SOC as well as the superconducting gap function~\cite{Brydon:2013aa,Schnyder:2013aa}. %
Moreover, SOC and relative amplitudes of singlet- and triplet-pairings in the superconducting gap can strongly affect the spin texture of the edge states in NCSs and eventually the transport properties at the surface~\cite{Tanaka:2009:PRB,Vorontsov:2008:PRL}.
Thus, the symmetries of superconducting gap and   SOC
strength support the different types of topological surface states 
 in noncentrosymmetric superconductors  like 
LiPt$_3$B~\cite{Togano:2004,Badica:2005,Yuan:2006,Nishiyama:2007}, 
CePt$_3$Si~\cite{Yogi:2004,Izawa:2005,Bonalde:2009}, 
CeRhSi$_3$~\cite{Kimura:2005}, CeIrSi$_3$~\cite{Sugitani:2006,Mukuda:2008} 
and 
Mo$_3$Al$_2$C~\cite{Bauer:2010,Karki:2010}. \\

Depending on the origin, two different types of antisymmetric SOC can be considered in noncentrosymmetric systems, which derived from bulk (\DsH)~\cite{Dresselhaus:1995} and structure (Rashba)~\cite{Rashba:1960} inversion asymmetry.
 The effects of the Rashba and \DsH~SOCs have been studied in ultra-cold fermions~\cite{Hui:2011,Yu:2011,DellAnna:2012,Li:2012,He:2012},
quantum wells~\cite{Ganichev:2004,Koralek:2009}, two-dimensional (2D) NSC systems~\cite{Yan:2014,Dias:2016},  Weyl semimetals~\cite{Volpez:2017}, 
transition metal dichalcogenides\cite{DiXiao:2012} (TMDs) and half-Heusler compounds\cite{Tafti:2013}.
Recently, the ultra cold fermions\cite{Sau:2011,Jiang:2011} and superconducting TMDs such as NbSe$_2$\cite{Hsu:2017,Xi:2015} have shown protected surface states in the presence of a magnetic field as a result of Rashba and/or~\DsH. 
 The \DsH~contribution has a cubic dependency in momentum while for Rashba it is linear.
Exceptionally, in quasi-two dimensional systems, due to the dimensional confinement and symmetry reduction, the leading term for \DsH~is linear in momentum\cite{SCHLIEMANN:2006aa}.
On top of that, regarding the point group symmetry of crystal, many restrictions are imposed to its mathematical form.
 The linear \DsH~SOC can be realized in the point groups such as $D_{2h}$~\cite{Winkler:2003,Ganichev:2014}.
 The symmetry reduction can be achieved by doping, temperature, oriented crystal growth and applying pressure.
 Spatial confinement generates a gradient in confining potential, which leads to induction the Rashba SOC.
Moreover, the Rashba term can be created by means of applying a normal electric field to a 2D surface~\cite{Vasco:1979,Bychkov:1984}. 
\\

In this paper, we investigate the topological properties of the 2D NCSs in the presence of Rashba and/or \DsH~antisymmetric spin-orbit couplings.
  We study the topological edge states using exact diagonalization of Bogoliubov-de~Gennes Hamiltonian on a ribbon geometry.
We analyze the response of the spin-polarized edge modes to the exchange field for a superconductor-ferromagnet~(FM) junction.
 Due to the coupling of spin-polarized edge states and exchange field, the edge modes may be chirally dispersing depending on the spin polarization direction. 
Therefore at the end, the response to the exchange field is studied as 
a nonchiral (uni-directioal) charge current at opposite edges.

 \begin{figure}[t]
 \begin{center}
\vspace{0.10cm}
 \hspace{-0.1cm}
\includegraphics[width=0.98 \linewidth]{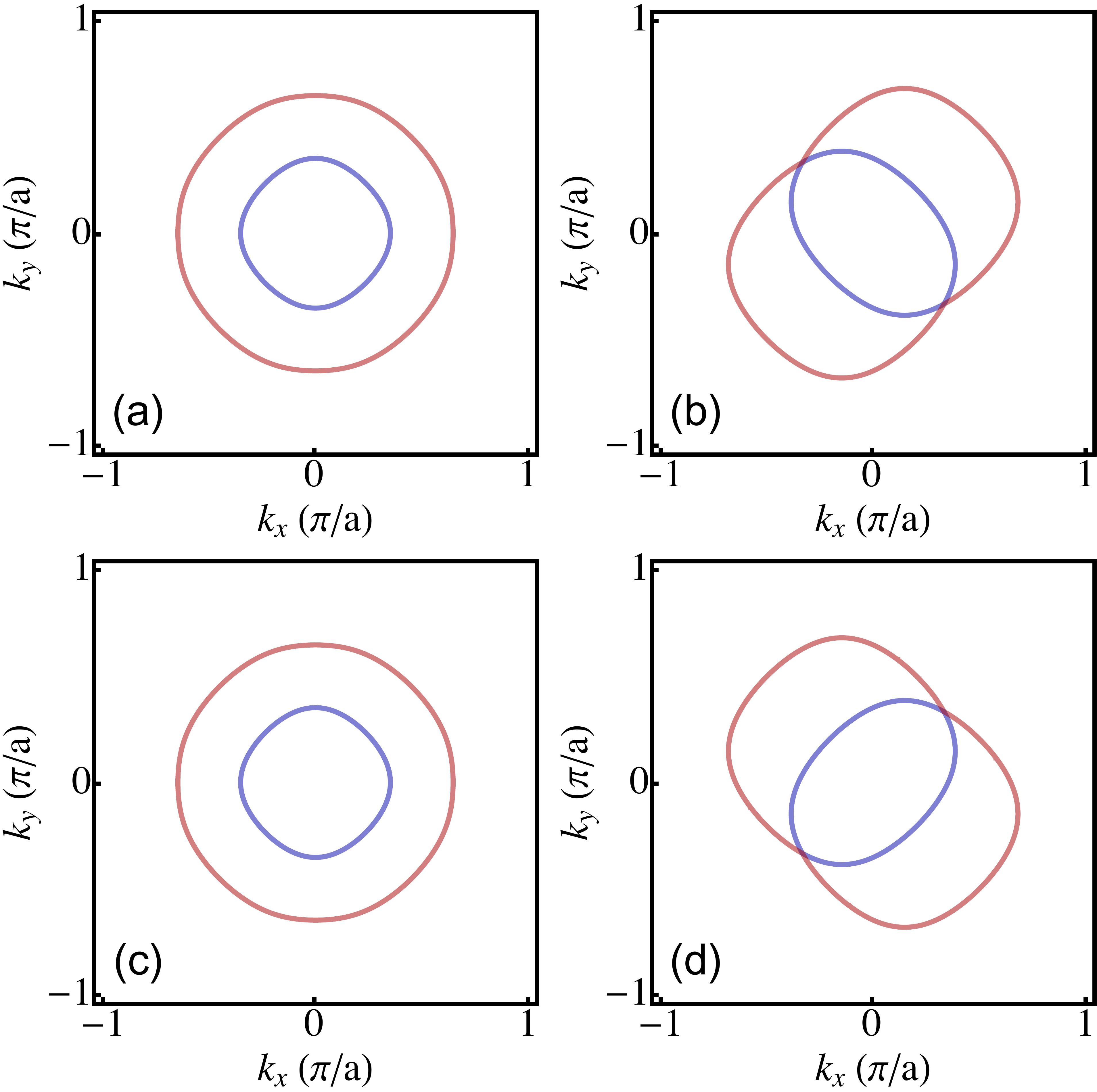} 
\end{center}
 \vspace{-0.2cm}
\caption{(Color online) 
 Fermi surfaces for a square lattice at the absence of an external magnetic field. The red line indicates the positive, $\varepsilon_{+}(\bk)=0$, and blue lines  represents the negative, $\varepsilon_{-}(\bk)=0$, helical Fermi surfaces. (a) Pure Rashba ($\alpha \neq 0, \beta=0$), (b) identical contributions of Rashba and \DsH~($\alpha=\beta=1/\sqrt{2}$), (c) pure \DsH~($\alpha=0, \beta \neq 0$), and opposite contributions of Rashba and \DsH~SOCs ($\alpha=-\beta=1/\sqrt{2}$). The spin split bands touch each other in a certain direction for $\alpha=\pm \beta$. 
}
\label{fig:Fig1}
\end{figure}

\section{Theoretical Model  Hamiltonian}
We consider a 2D single band superconductor  without an inversion symmetry on the square lattice (lattice parameter set  to $1$) and with the Hamiltonian given by
\be
{\cal H}=\frac{1}{2}\sum_{\bk}\Psi^{\dagger}_{\bk}{\cal H}(\bk)\Psi^{}_{\bk},
\ee
where  
$\Psi^{\dagger}_{\bk}=(c^{\dagger}_{\bk,\uparrow},c^{\dagger}_{\bk,\downarrow},c^{}_{-\bk,\uparrow},c^{}_{-\bk,\downarrow})$
is  the Nambu space operator, and 
\be
 {\cal H}({\bk}) = \begin{pmatrix} \hh(\bk)  & \Delta(\bk)  \\
 \Delta^{\dagger}(\bk)  &     -\hh^{*}(-\bk)
\end{pmatrix},
\label{Eq:Bulk_Ham}
\ee
%
 where
 $c^{\dagger}_{\bk,\sigma}$ ($c^{}_{\bk,\sigma}$) creates (annihilates) an electron with momentum $\bk=(k_x,k_y)$ and spin $\sigma$, 
 and $\Delta(\bk)$ is the superconducting  matrix gap function in spin basis.
Furthermore, the normal state Hamiltonian, $\hh(\bk)$, is described by the $2 \times 2$ matrix
%
\be
\hh(\bk)=\epsilon(\bk) \sigma_0 +{\bg}(\bk)\cdot {\bm \sigma}-\mu_B \bH \cdot {\bm \sigma},
\label{Ham_matrix}
\ee
%
 Assuming a nearest-neighbor tight binding model with a hopping $t$, for the conduction electron bands with a total band width $W=8t$ and  the chemical potential $\mu$, the dispersion is given by
$\epsilon(\bk)=-2t(\cos{k_x}+\cos{k_y})- \mu$.
 Finally  the  effective field $\bH=(H_x,H_y,H_z)$ is  introduced with a Zeeman term, ${\cal H}_Z=-\mu_B \bH \cdot {\bm \sigma}$.\\

The broken spatial inversion symmetry induces an antisymmetric SOC originated from bulk or structure inversion asymmetries (BIA and SIA).
Microscopically, BIA  results from  the absence of  inversion symmetry in the bulk of  the material and leads to the \DsH~SOC.
 However, SIA originates from the inversion asymmetry of the
confining potential and generates the Rashba term in the Hamiltonian whose strength can be manipulated by an external
electric field.
In the mixed condition, fine tuning of Rashba and \DsH~SOCs is possible through external gates or doping profile~\cite{Engels:1997,Nitta:1997}. 
These SOCs are defined by a characteristic $\bg$-vector 
\be 
{\bg}(\bk)=\alpha {\bg^{R}(\bk)}+\beta  {\bg^D(\bk)},
\ee
%
 which is an odd function with respect to $\bk$
 $$ \bg(-\bk)=-\bg(\bk),  $$
  and the Rashba and \DsH~$\bg$-vectors defined  by 
\bea
\no
{\bg^{R}(\bk)}={\rm g}(\sin{k_y},-\sin{k_x});
\\\no
{\bg^D(\bk)}={\rm g}(\sin{k_x},-\sin{k_y}).
\eea
 Here,   ${\rm g}$ is the SOC magnitude, and
 the control parameters   
 $ |\alpha|, |\beta| \in [0,1]$, 
 tune  the magnitudes of Rashba and \DsH~SOCs. 

Antisymmetric SOC, playing the role of a $\bk$-dependent magnetic field, leads to a locking of the spin and orbital degrees of freedom.
 This results in lifting the two-fold spin degeneracy and  splits the Fermi surface into two opposite-helicity sheets.
 In the  diagonal helicity basis, the normal Hamiltonian is
%
\be
\tilde{h}(\bk)=
\begin{pmatrix}
\varepsilon_{+}(\bk) & 0 
\\
0 &\varepsilon_{-}(\bk)
\end{pmatrix}
,
\ee
with the following eigenvalues
\begin{align}
\begin{aligned}
&
\varepsilon_{\pm}(\bk)
=
\epsilon(\bk)  
\\
& \pm
\sqrt{
\mu_B^2 H_z^2+
\Big(
{\rm g}_x(\bk)-\mu_B H_x
\Big)^2+
\Big(
{\rm g}_y(\bk)-\mu_B H_y
\Big)^2}.
\end{aligned}
\end{align}
%
%
Therefore, in the absence of Zeeman field, the eigenvalues are given by
%
%
\begin{align}
\begin{aligned}
&
\varepsilon_{\pm}(\bk)
=
\epsilon(\bk) 
\\
& 
\pm
{\rm g} \sqrt{(\alpha^2+\beta^2)(\sin^2{k_x}+\sin^2{k_y}) +4\alpha\beta \sin{k_x} \sin{k_y}}~.
\end{aligned}
\label{Eq:Normal_energy_spec}
\end{align}
%
Fig.\ref{fig:Fig1} displays the evolution of Fermi surface for the different  strength of Rashba and \DsH.
 For pure Rashba (\DsH), Fermi surface has an isotropic structure in the Brillouin zone (BZ) with
 $C_{4v}$
  point group symmetry.
  The simultaneous presence of both SOCs leads to reduction of symmetry group to
  $C_{2v}$   
   with an anisotropic structure of Fermi surfaces~\cite{Yan:2014}.
   In two special cases with $\alpha=\pm \beta$, the SOC split bands touch each other at certain directions $[1 \bar{1}]$ and $[1 1]$ with special consequences on topology of the electronic band structure, which will be discussed later.

Due to the broken inversion symmetry the  parity is not a well-defined quantum number, therefore  the gap function, $\Delta({\bk})$, has to include  both  singlet (even-parity) and triplet  (odd-parity) components~\cite{Frigeri:2004}, simultaneously, i.e. 
$$\Delta(\bk)=(\psi_{\bk}\sigma_0+\bd_{\bk}\cdot{\bm \sigma})({\mi}\sigma_y),$$
where the spin-singlet and spin-triplet components of the superconducting order parameter are described by 
\bea
\no
\psi_{\bk}
&=&
r\Delta_0 f(\bk),
\\
 \bd_{\bk}
&=&
(1-r)\Delta_0 f(\bk) \hat{\bg}(\bk),
\eea
respectively.
 Here $\Delta_0$ is the superconducting pairing amplitude,   and the dimensionless parameter $r$  varies from $0$ (pure triplet) to $1$ (pure singlet) determining  the dominant component in the superconducting state, 
 and   $\sigma_i$ ($i=x,y,z$) are the Pauli matrices in spin space.
 In the absence of the  inversion symmetry 
  the triplet component survives, provided that $\bd_{\bk}$ is aligned with the SOC $\bg$-vector, i.e. $\bd_{\bk}||\bg_{\bk}$\cite{Frigeri:2004}. 
  Moreover, the structure factor $f(\bk)$ is introduced to describe possible higher orbital angular momentum pairing
     for the superconducting gap.
   For a 2D system with 
    $C_{4v}$
    point group symmetry, the allowed irreducible representations are $A_1$, $B_1$ and $B_2$~\cite{Dresselhaus:2008aa}.
    Thus, only the singlet form factors with $k_xk_y$ and $k_x^2 \pm k_y^2$ symmetry are permitted to occur.
 In the following, we  study the topology of electronic structures in prototypical noncentrosymmetric superconductors with (translationally invariant)  structure factors
\be
f(\bk)=\begin{cases}
1 & \text{ A$_1$ ($s$-wave)}
\\
\sin{k_x} \sin{k_y} & \text{ B$_2$ ($d_{xy}$-wave)}
\end{cases}.
\label{Eq:Structure_factor}
\ee
%
%
%
 Because $\bd_{\bk}||\bg_{\bk}$ the  odd parity triplet part corresponds to $p$-wave pairing and then the mixed 
singlet triplet gap functions $\Delta(\bk)$ are either of $s+p$ or $d_{xy}+p$ type.  Note that the triplet part of the latter has only two sign changes on each of the split Fermi surface sheets,
therefore the designation $p$-wave  may be used also in this case~\cite{Tanaka:2010aa}.
 \begin{figure*}[t]
 \begin{center}
 \hspace{-0.1cm}
 \includegraphics[width=1 \linewidth]{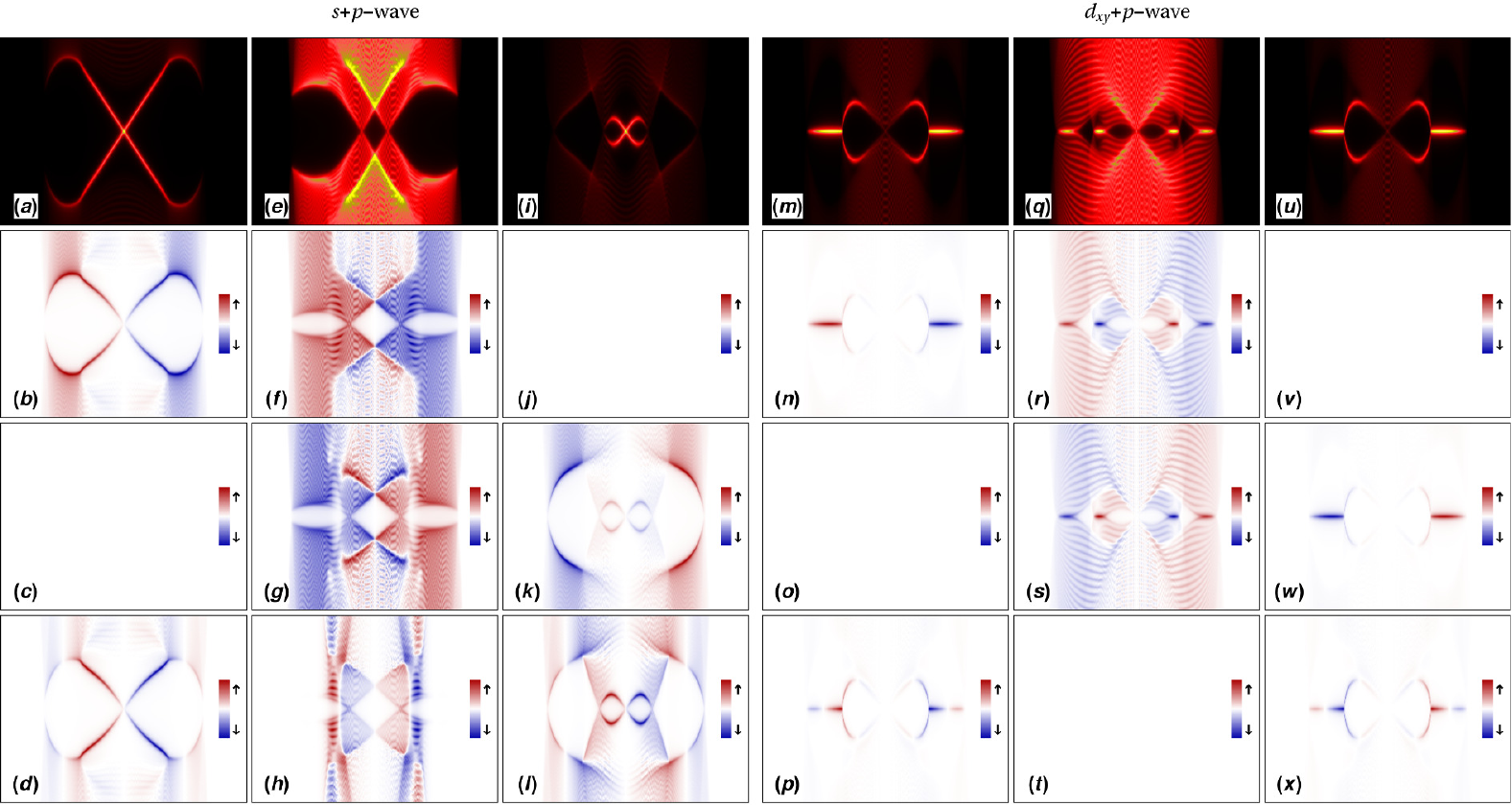} 
\end{center}
\caption{(Color online) 
Energy dispersion of the
 momentum-resolved and spin-resolved spectral functions for a triplet-dominant noncentrosymmetric superconductor ($r=0.25$);
Left panel represents the $s+p$, and right panel displays the $d_{xy}+p$ pairings.
The first row shows the momentum-resolved spectral function, while the second to fourth rows denote $x$-, $y$-, and $z$-components of spin-resolved spectral function, respectively.
At each panel, the first, second and third columns exhibit the  pure Rashba ($\alpha=1,\beta=0$), mixed Rashba and \DsH~($|\alpha|=|\beta|=1/\sqrt{2}$), and pure \DsH~($\alpha=0,\beta=1$), respectively.
 Notes: Pure Rashba has no $y$-component, (c and o), and  pure \DsH has no $x$-component, (j and v),  of spin-polarizations. For $d_{xy}+p$-wave with mixed Rashba-\DsH the spin-polarization vector is in $xy$-plane, (t).
  In all sub-plots: The vertical axes refer to quasi-particle energy,~$-t\le \omega\le t$,
   and
  the axes of abscissa represent the    $k_y\in [-\pi/a,\pi/a]$.
}
\vspace{0.73cm}
\label{fig:Fig2}
\end{figure*}
%
%
 \begin{figure*}[t]
 \begin{center}
 \hspace{-0.1cm}
 \includegraphics[width=1 \linewidth]{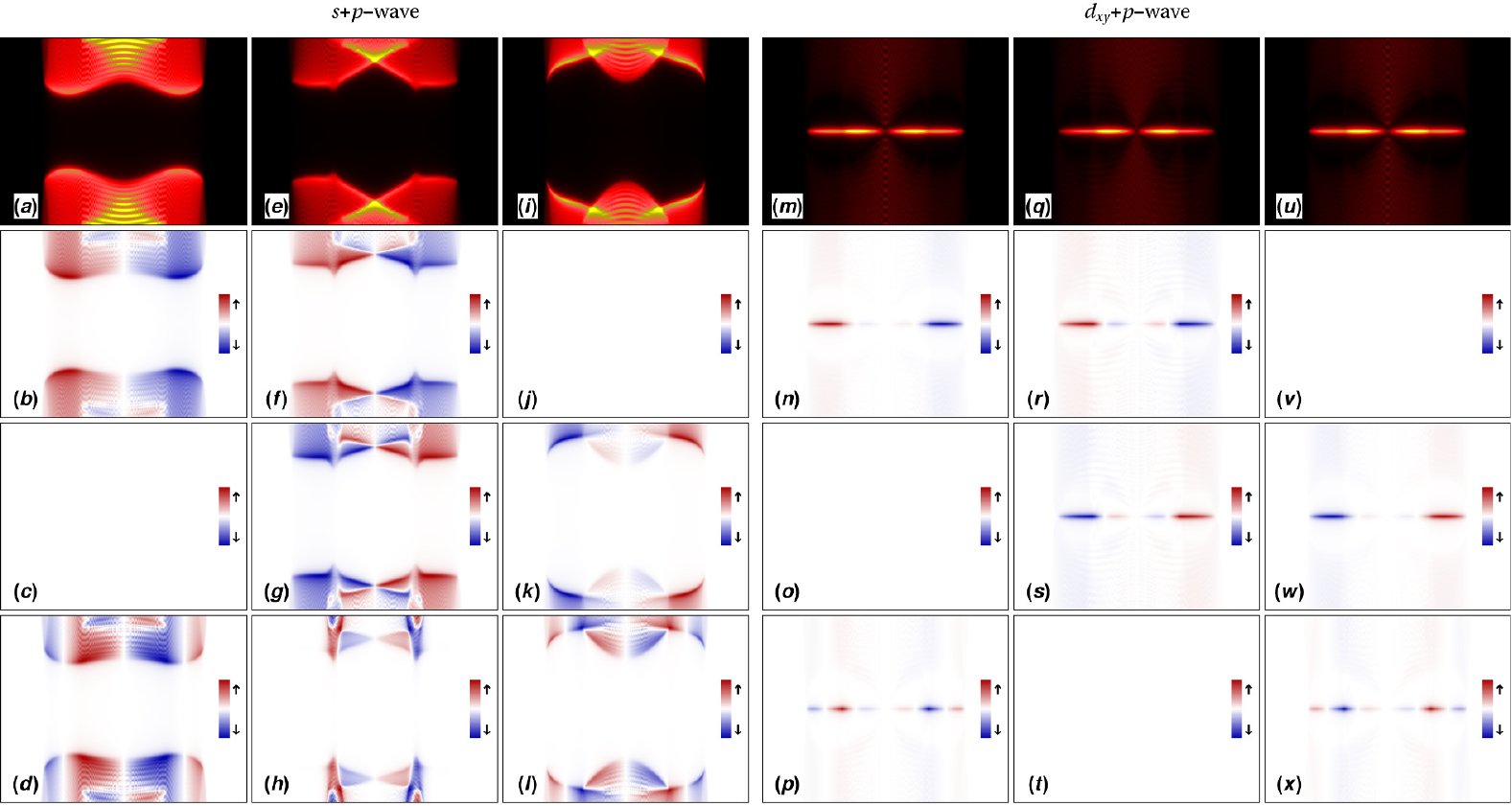} 
\end{center}
\caption{(Color online)
Same as Fig.~\ref{fig:Fig2}  except that it is depicted for the singlet-dominant pairing ($r=0.75$). Notes: Compared to Fig.~\ref{fig:Fig2}, the singlet-dominant $s+p$-wave shows a trivial topology.
}
\label{fig:Fig3}
\vspace{0.6cm}
\end{figure*}
%
\section{Topological Properties}
In order to study the topology of the electronic band structure and edge modes of a two-dimensional  NCS, one has to examine the existence of ingap states.
 For this purpose, the model Hamiltonian   is diagonalized  on a ribbon geometry with cylindrical boundary condition, i.e. open and periodic boundaries along $x$- and $y$-directions, respectively.  
 In this configuration, we suppose the 2D system as $N_x$ parallel  rods     along the $y$-direction.
 Therefore, the mapping of bulk BZ into rod geometry formalism can be done by a partial Fourier transform of electron creation operator, 
 $$
 c^{\dagger}_{\bk,\sigma}=\frac{1}{N_x}\sum_{n}e^{{\mi} k_x x_n} c^{\dagger}_{n,k_y,\sigma}.
 $$
 Here $c^{\dagger}_{n,k_y,\sigma}$ creates an electron in $n$th rod   ($n=1,2,...,N_x$) with momentum $k_y$ and spin $\sigma$.\\

 The Bogoliubov-de Gennes Hamiltonian within the generalized Nambu space $\Psi^{\dagger}_{k_y}=(\Phi^{\dagger}_{1,k_y},\Phi^{\dagger}_{2,k_y},...,\Phi^{\dagger}_{N_x,k_y})$  is given by
%
\be
{\cal H}=\frac{1}{2}\sum_{k_y}\Psi^{\dagger}_{k_y}{\cal H}_{BdG}(k_y)\Psi^{}_{k_y},
\ee
where  $\Phi^{\dagger}_{n,k_y}=(c^{\dagger}_{n,k_y,\uparrow},c^{\dagger}_{n,k_y,\downarrow},c^{}_{n,-k_y,\uparrow},c^{}_{n,-k_y,\downarrow})$.
Here the matrix form of Bogoliubov-de Gennes Hamiltonian including $4 \times 4$ matrices $M$, $T_1$, and $T_2$ is obtained by
%
%
\be
{\cal H}_{BdG}(k_y)=
\begin{bmatrix}
M & T_1 & T_2 &  0&   &   & ...
\\
T^\dagger_1 & M & T_1 & T_2 & 0  &   & ...
\\
T^\dagger_2 & T^\dagger_1 & M & T_1 & T_2 &  0 & ...
\\
 0 & T^\dagger_2 & T^\dagger_1 & M & T_1 & T_2 &  ...
\\
 & & & & &  & \ddots
\end{bmatrix}.
\label{Eq:Slab_Ham}
\ee
%
%
The matrix $M$ stands for intra-rod hopping as well as on-site energies, however the matrices $T_1$ and $T_2$ describe the inter-rod hopping between the nearest and the next nearest rods, respectively. 
These are satisfying the following recursive equation on the basis of $\Phi^{\dag}_{n,k_y}$
%
%
\begin{align}
\begin{aligned}
&T^{}_2 \Phi_{n-2,k_y,\sigma}
 +
T^{}_1 \Phi_{n-1,k_y,\sigma} +
M \Phi_{n,k_y,\sigma} 
\\
& +
T^{\dag}_1 \Phi_{n-1,k_y,\sigma} +
T^{\dag}_2 \Phi_{n-2,k_y,\sigma} =
\zeta_n
 \Phi_{n,k_y,\sigma}
\label{Eq:resursive}
\end{aligned}
\end{align}
%
%
with $\Phi_{0,k_y,\sigma}=\Phi_{N_x+1,k_y,\sigma}=0$ as the boundary condition. Here $\zeta_n$ is a $4 \times 4$ diagonal matrix, whose elements give the energy spectrum of the system in the $n$th rod.
 Among the solutions of the recursive relation, we are interested in the solutions exponentially decaying along $x$-direction, which manifest the existence of nontrivial topological edge states. 
The definitions of $M$, $T_1$ and $T_2$ matrices  depend on the symmetry of superconducting gap function. 
Following the discussions before Eq.~(\ref{Eq:Structure_factor}), for an $s+p$-wave superconductor, the intra-rod energy is
%
\begin{align}
\begin{aligned}
M
 =&
-\Big( \mu+2t\cos{k_y} \Big) \tau_z
+
{\rm g}
\sin{k_y}
 \Big( \alpha  \sigma_x- \beta  \tau_z \sigma_y \Big) 
\\
&
-r \Delta_0
 \tau_y \sigma_y-
 (1-r) {\Delta_0}
 \sin{k_y}
  \Big( \alpha  \tau_x \sigma_z - \beta  \tau_y \Big) ,
\end{aligned}\no
\end{align}
%
%
%
whereas the inter-rod hopping are given by
%
%
\begin{align}
\begin{aligned}
T_1
=& -t  \tau_z 
+ \frac{{\mi}}{2}
 {\rm g}
  \Big( \alpha \tau_z \sigma_y 
-
\beta  \tau_z \sigma_x \Big)
\\&
 -{\mi} 
 (1-r)
   {\Delta_0} 
 \Big( \alpha \tau_y - \beta  \tau_x \sigma_z \Big),
\end{aligned}\no
\end{align}
%
%
%
 and $T_2=0$.
Furthermore, in a superconducting system with $d_{xy}+p$-wave Cooper pairing, we can find
%
%
\begin{align}
\begin{aligned}\no
M 
=&
-\Big( \mu+2t\cos{k_y} \Big) \tau_z+{\rm g} \sin{k_y} \Big( \alpha \sigma_x - \beta \tau_z\sigma_y \Big) 
\\
&
+\frac{1}{2} (1-r) {\Delta_0} \sin{k_y} \Big( \alpha \tau_y - \beta \tau_x \sigma_z \Big),
\end{aligned}
\end{align}
%
%
%
%
%
\begin{align}
\begin{aligned}\no
T_1
=
&-t\tau_z +\frac{{\mi}}{2} {\rm g} \Big( \alpha \tau_z \sigma_y - \beta  \sigma_x \Big)+ \frac{\mi}{2} r {\Delta_0} \sin{k_y} \tau_y \sigma_y
\\
&
+\frac{\mi}{2} (1-r){\Delta_0} \sin^2{k_y} \Big( \alpha \tau_x \sigma_z - \beta \tau_y \Big),
\end{aligned}
\end{align}
%
%
%
and
%
%
\bea
\hspace{-1.5cm}
T_2=-\frac{1}{4}(1-r) {\Delta_0}  \sin{k_y} \Big( \alpha \tau_y - \beta \tau_x \sigma_z \Big),
\no
\eea
where $\tau_i$  ($i=x,y,z$) are the Pauli matrices in the particle-hole  space. 
 In our calculations, we set the values of parameters as 
$t=2$, $\mu=4$, ${\rm g}=2$,  and $\Delta_0=0.5$. The latter are still small against the total electronic bandwidth $W=8t$: $g/W=1/8$ and $\Delta_0/W=1/32$.
\\

By use of the Matsubara and retarded Green's functions for the ribbon geometry as 
$$\hat{G}(k_y,{\mi} \omega)=[{\mi} \omega-H(k_y)]^{-1},$$
 and 
 $$\hat{G}^{ret}(k_y,\omega)=
 \hat{G}(k_y,{\mi} \omega)
 |_{ {\mi}\omega \rightarrow \omega+{\mi} 0^{+}},$$
  one can define the local density of states (LDOS) for the $n$th rod as 
%
%
\be
N_n(\omega)=-\frac{1}{\pi}\sum_{k_y}{\rm Im} \Big[ {\rm Tr}_{} \Big\lbrace G_{nn}^{ret}(k_y,\omega) \Big\rbrace \Big].
\label{Eq:Slab_LDOS}
\ee
%
Calculation of LDOS in the triplet-dominant case ($r=0.25$)  shows that both symmetries  realize nontrivially topological edge states, 
however the $s+p$-wave represents a trivial topology for the singlet-dominant  ($r=0.75$), compatible with Ref.~[\onlinecite{Sato_Fujimoto:2009}].

%

 \begin{figure}[t]
 \begin{center}
\vspace{0.4cm}
 \hspace{-0.2cm}
 \includegraphics[width=0.95 \linewidth]{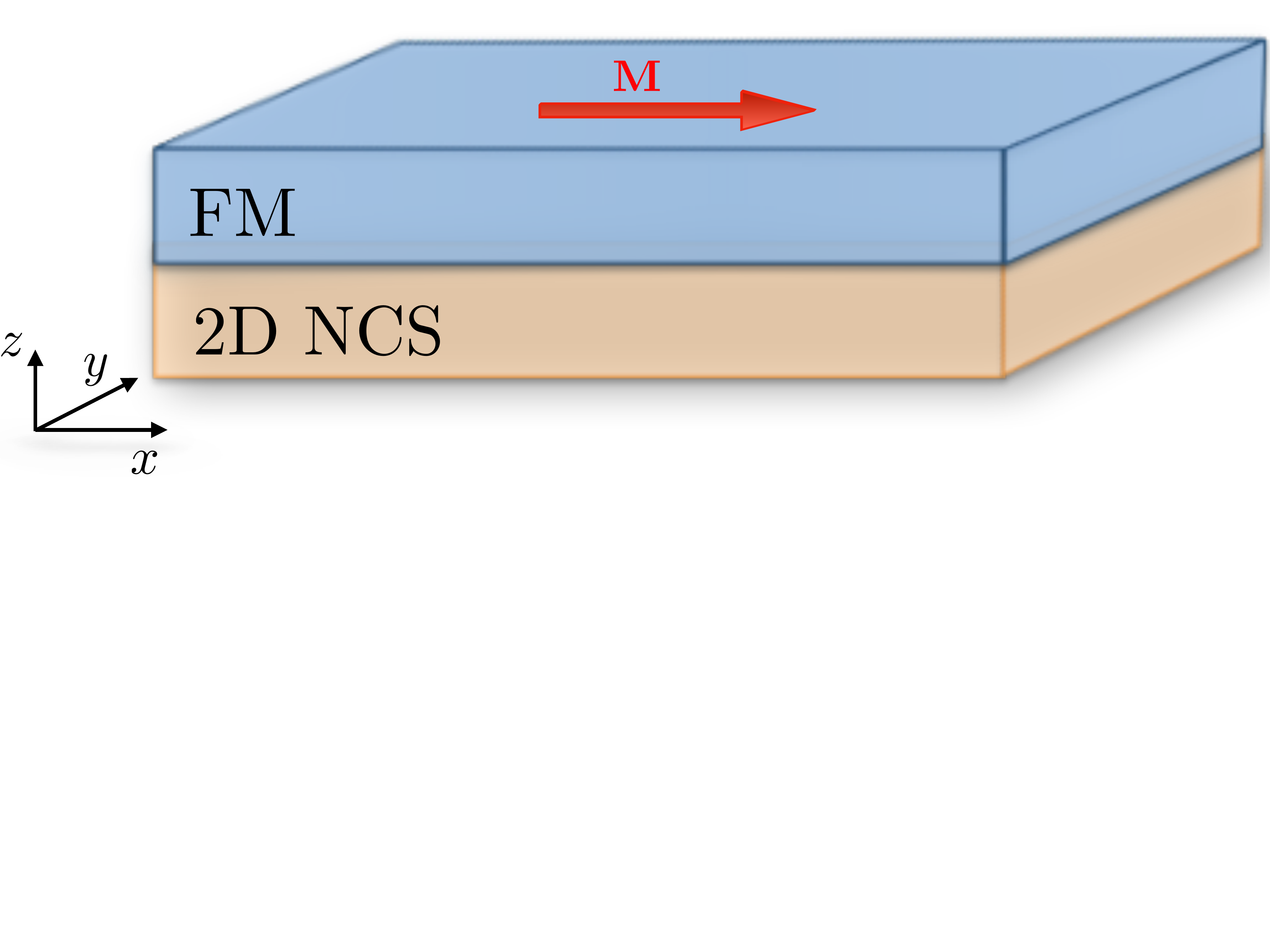} 
\end{center}
 \vspace{-3.4cm}
\caption{(Color online)
 Schematic diagram of the NCS-FM heterostructure considered in this work. The NCS and FM occupy the $z<0$ and $z>0$ half spaces, respectively. The FM magnetization, $\bM$,  points in $x$-direction.
 Then ${\bf H}=(z|J|/g^2\mu_B^2){\bf M}$ is the effective exchange field  in Eq.~(\ref{Ham_matrix}) generated by the FM magnetization {\bf M} (per site) at the FM/NCS interface,
here $J=$ FM-exchange constant, $g=$ g-factor of ordered moments, and $z=$ coordination number in FM. 
To avoid the electron tunneling between NCS and FM a very thin insulating barrier is placed at $z=0$ or a bulk insulating FM phase is used. 
}
\label{fig:Fig4}
\end{figure}
%

%
\begin{SCfigure*}
  \centering
 \includegraphics[width=1.4 \linewidth]{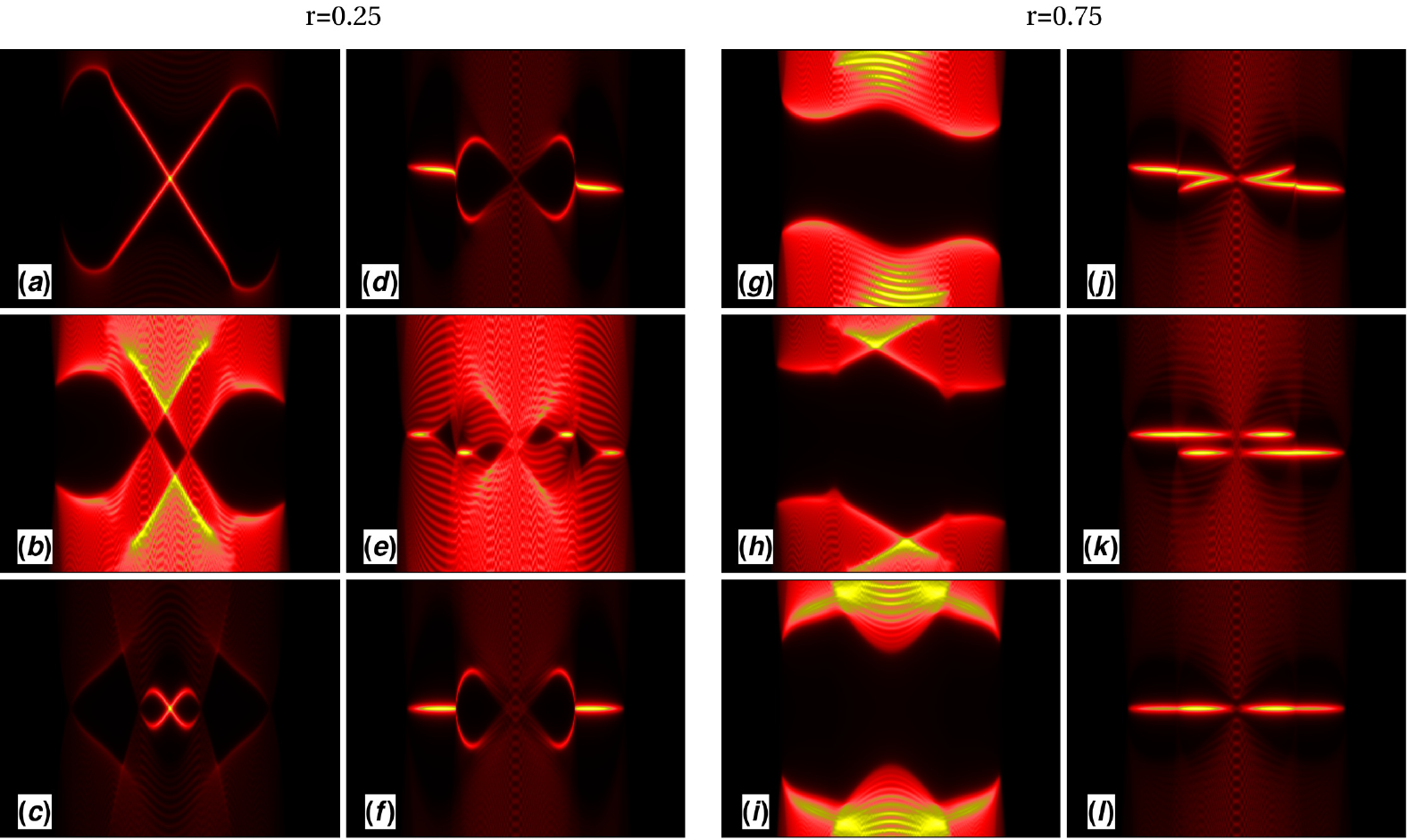} 
   \caption{ (Color online)
   Energy dispersion of the
   momentum-resolved spectral function for the edge rod of an NCS  in the presence of a exchange field along $x$-direction with $\mu_{B}H_x=0.05$;
    left panel represents a  triplet-dominant ($r=0.25$), and right panel shows a singlet-dominant ($r=0.75$).
At each panel, the left and right columns denote the $s+p$- and $d_{xy}+p$-pairings, respectively. 
The first, second and third rows represent the  pure Rashba ($\alpha=1,\beta=0$), mixed Rashba and \DsH~($|\alpha|=|\beta|=1/\sqrt{2}$), and pure \DsH~($\alpha=0,\beta=1$), respectively. 
 In all sub-plots: The vertical axes refer to quasi-particle energy,~$-t\le \omega\le t$,
   and
  the axes of abscissa represent the    $k_y\in [-\pi/a,\pi/a]$.
   \vspace{0.cm}
  }
  \vspace{0.4cm}
  \label{fig:Fig5}
\end{SCfigure*}
%
%

%
To consider the realization of nontrivial topology in the band structure on an NCS, we define the momentum resolved spectral function, which for $n$th rod is given by
%
\be
\rho_n(k_y,\omega)=-\frac{1}{\pi} {\rm Im} \Big[ {\rm Tr} \Big\lbrace  G_{n}^{ret}(k_y,\omega) \Big\rbrace \Big].
\label{Eq:Momentum_Resolved_LDOS}
\ee
%
%
Due to the scattering of electron spin by SOC, the topological  modes  are strongly spin-polarized.
 Therefore, to study the spin texture of the topological edge states, we obtain  the spin-resolved spectral function for $n$th rod  as
%
%
\be
\rho^{\nu}_n(k_y,\omega)=-\frac{1}{\pi} {\rm Im} \Big[ {\rm Tr} \Big\lbrace \check{S}_{\nu} G_{n}^{ret}(k_y,\omega) \Big\rbrace \Big],
\label{Eq:Spin_Resolved_LDOS}
\ee
%
%
%
where 
$$
\check{S}_{\nu}=\frac{1}{2} 
\Big[ 
(1+\tau_z) \hat{\sigma}_{\nu}
-
(1-\tau_z) \hat{\sigma}^{*}_{\nu}
\Big],
$$
 and 
 $\nu=(x,y,z)$.
 The traces in the Eqs.~(\ref{Eq:Slab_LDOS}-\ref{Eq:Spin_Resolved_LDOS})
perform over particle-hole and   spin  spaces.
Fig.~\ref{fig:Fig2} shows the momentum- and spin-resolved spectral functions for a triplet-dominant  
 $s+p$- (left panel), and $d_{xy}+p$-wave  (right panel) NCS.
 Figs.~\ref{fig:Fig2}(a~\&~m)  present 
a system with pure Rashba SOC ($\alpha=1,~\beta=0$) for $s+p$- and $d_{xy}+p$-pairings, respectively.
 In addition, Figs.~\ref{fig:Fig2}(e~\&~q) and Figs.~\ref{fig:Fig2}(i~\&~u) correspond to a NCS with the interplay of Rashba and \DsH~with the same strength ($\alpha=\pm\beta=1/\sqrt{2}$), and with pure \DsH~($\alpha=0,~\beta=1$), respectively. 
 These results obviously reflect the consequences of superconducting gap  on the type of ingap states, which belong to DIII class of topological superconductors~\cite{Schnyder:2008,Schnyder:2016}. 
 Due to the nodal structure of $d_{xy}+p$-wave state, the zero-energy edge modes appear as flat-bands.
 Note that none of the global topological invariants characterize these flat bands, in contrast to the $s+p$-wave state described by the global $\mathbb{Z}_2$ number.
All of these states are topologically non-trivial and potentially host Majorana fermions, except  the case of a $s+p$-wave NCS with the same contributions of Rashba and \DsH. 
As it can be seen in Fig.~\ref{fig:Fig1}(b~\&~d), two helical bands touch each other at certain directions for $|\alpha|=|\beta|$. 
This is equivalent with the local disappearance of SOC $\bg$-vector. 
Since the triplet component, 
 $\bd_{\bk}$,
  is
tied to the existence of
 $\bg_{}(\bk)$,
the superconducting gap function should be  fully s-wave with  trivial topology
 for special points on the Fermi surface  that justify the condition of $\sin k_x=-\sin k_y$. 

First row in Fig.~\ref{fig:Fig3} shows the momentum-resolved spectral function at the surface of a singlet-dominant  
 $s+p$-wave (a,~e,~and~i), and $d_{xy}+p$-wave (m,~q~and~u)   NCS.
 Because of fully gapped structure, the  singlet-dominant $s+p$-wave NCS does not show nontrivial topology. 
Second to fourth rows of Fig.~\ref{fig:Fig2}~and~Fig.~\ref{fig:Fig3} represent the $x$-,~$y$-,~and~$z$-~components of spin-resolved spectral functions in NCSs.
 One can easily observe that the existence of  strong SOC causes both electron- and hole-like parts of the edge states exhibiting a distinct spin texture.
  Therefore, as a result of   
  time-reversal symmetry   two counter-propagating  modes flow  at each edge with opposite spin polarization. 
Figs.~\ref{fig:Fig2}~and ~\ref{fig:Fig3} show that for both pairing symmetries the spin-polarization vectors are in  $xz$-~($yz$-)~plane for pure Rashba (\DsH).
In a $s+p$-wave NCS with $|\alpha|=|\beta|$, the spin polarization vector has all $x$-,~$y$- and $z$-~components, while the $d_{xy}+p$-wave one has only $x$- and $y$-elements.

\section{ Nonchiral edge currents}
In the presence of time-reversal symmetry, there are two  helical  modes with opposite spin polarizations at each edge of NCSs. 
These  modes propagate in opposite directions with a zero total  charge current.
Broken time-reversal symmetry together with a slight shift in the energy 
 of edge state result in generation of 
 a net charge current at the edge. 
In this section,  we investigate the edge modes at the interface of an NCS and a ferromagnet (NCS-FM) junction.
In a ferromagnet, the spins of individual atoms are coupled by the direct- or super-exchange  interactions. 
The resulting net magnetic moment oriented along the easy direction leads to an effective exchange field [${\bf H}$ in Eq.~(\ref{Ham_matrix})] in the ordered phase.
  At an NCS-FM junction it also modifies  the energy of  spin-polarized  states on the NCS side.
Schematically,    Fig.~\ref{fig:Fig4} shows a planar junction between the 2D NCS and the ferromagnet that are separated by a very thin insulating barrier to avoid electron tunneling between the superconducting and ferromagnetic environments. 
 Equivalently the FM bulk may be assumed to be an insulator.
 We restrict the current study only for the in plane magnetization because of completely different topological behaviors in $z$-direction~\cite{Sato_Fujimoto:2009}.\\
%

 \begin{figure}[t]
 \begin{center}
\vspace{0.cm}
 \hspace{-0.1cm}
 \includegraphics[width=0.99 \linewidth]{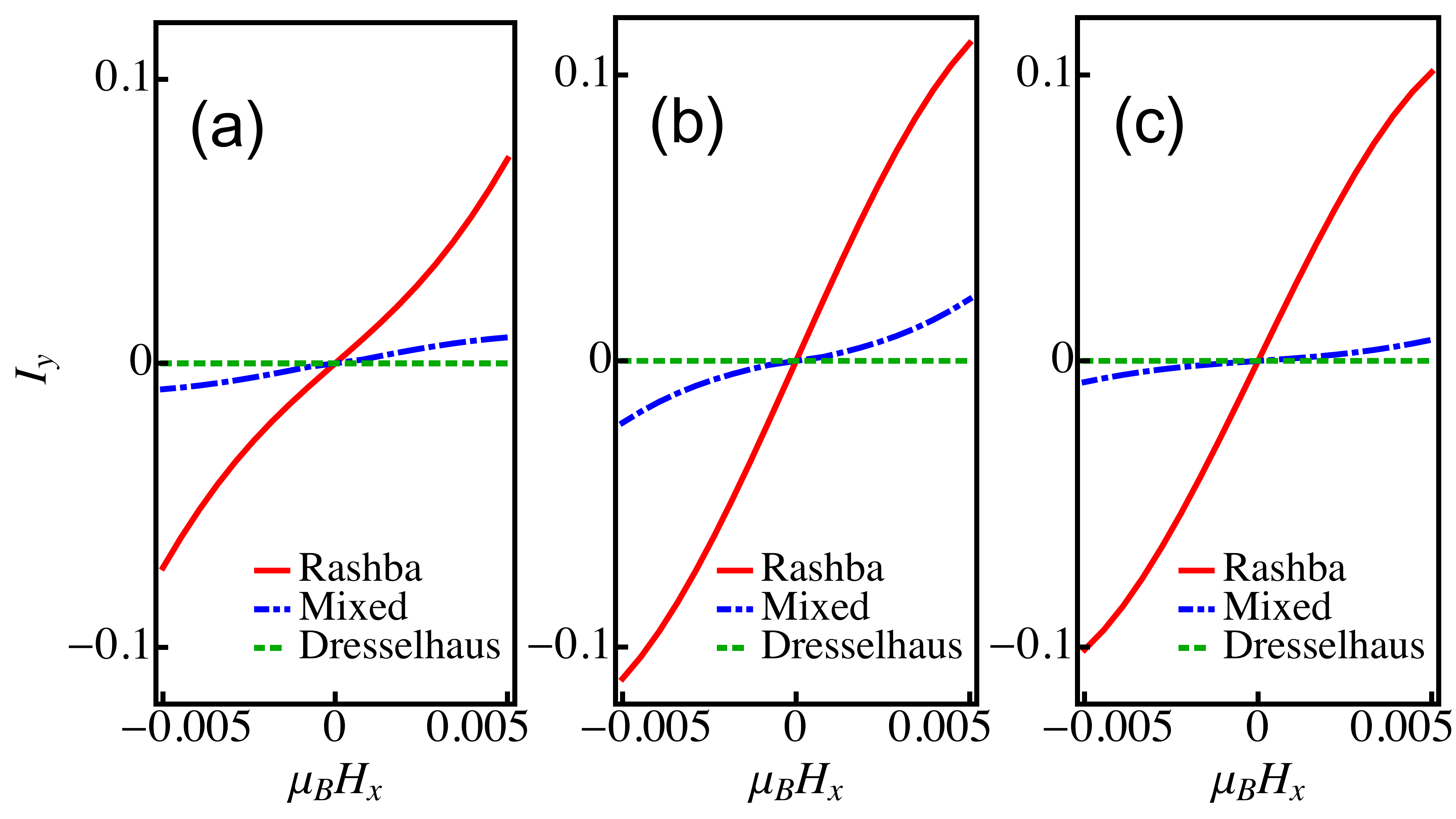} 
\end{center}
 \vspace{-0.3cm}
\caption{(Color online) 
Topological current at interface of NCS-FM junction for: 
(a) triplet-dominant  ($r=0.25$) with $s+p$-wave, (b)  triplet-dominant   ($r=0.25$) with  $d_{xy}+p$-wave, and (c) singlet-dominant ($r=0.75$) with $d_{xy}+p$-wave superconducting states.
The current unit is given by $et/(2\hbar )$.
}
\label{fig:Fig6}
\end{figure}
%

Fig.~\ref{fig:Fig5} depicts the momentum-resolved spectral functions for the $s+p$-wave and $d_{xy}+p$-wave NCSs in the presence of an external magnetic (exchange) field in $x$-direction with $\mu_B H_x=0.005$. 
 With  $\mu_B H_x/\Delta_0=0.001$ this is still far below the Pauli limiting field $0.7\Delta_0$ of the singlet component so that the superconducting state may be assumed as unaffected.
 Energies of 
the left- and right-moving edge states become different for $s+p$-wave pairing, and it leads to a   chirally dispersed  band structure, in contrast to helical modes in Fig.~\ref{fig:Fig2}.
On top of this,  the spin-polarized flat-bands shift in opposite direction and give a net charge current for the $d_{xy}+p$-pairing [see~Fig.~\ref{fig:Fig5}(d,e,j,k)]. 
 One can easily observe that this energy modification  shifts  all of the edge states  except those  
 with pure \DsH.
  This can be justified by the fact that the slight shift in the energy is the result of coupling of the exchange field and the spin-polarization at the edge.
The edge modes are not influenced by $H_x$, hence  $x$-component of the spin-polarization  is absent for  pure \DsH.
 Therefore we do not expect to detect non-zero 
  charge current in an NCS with pure \DsH.
Similarly,  
pure \DsH~shows a considerable contribution in 
  charge current for the exchange field along y-direction, while pure Rashba  generate no current.\\

In presence of exchange field the helical edge modes acquire different velocities that generate a non-zero current density. For $n$th rod, the current density for unit area is given by
  \be
  I_y^n=-e \langle v^n_y  \rangle, 
  \ee
 with velocity operator $v^n_y=\partial \hat{h}(\bk)/ \partial k_y$ described by
%
%
\begin{align}
\begin{aligned}
&
v^n_y 
=
\\
&
- 
\!\!
\Big[
 2t \sin k_y \tau_z 
-
{\rm g}\cos k_y (\alpha \sigma_x - \beta \tau_z \sigma_y)  
\Big]_{\sigma \sigma^{\prime}}
 \!
  c^{\dag}_{n,k_y,\sigma} c^{}_{n,k_y,\sigma},
\end{aligned}
\label{Eq:Velocity_operator}
\end{align}
%
%
where $\bra \cdots \ket$ defines an averaging over all quantum states at zero temperature,
 and $e$  is the electron charge.
The expectation value of the velocity operator   is given by
\be
\langle v^n_y \rangle = \frac{1}{N_y} \sum_{k_y} \int_{-\infty}^{0} d\omega~ {\rm Tr}_{} 
\Big\{
v^n_y 
\Big\},
\ee
where $N_y$ is the number of $k_y$ points in the edge Brillouin zone.
Since the  energy spectrum of Bogoliubov quasiparticles is always particle-hole symmetric around $\mu$,
then only states with $\omega<0$ contribute to the current and the integration over frequency is performed for all filled states. 

 The total topological current is obtained by doing the summation over contributions from all rods, i.e.
%
%
\begin{align}
\begin{aligned}
I_y
&
\!=\!
\sum_{n=1}^{N_x/2}
\sum_{k_y}
\frac{e}{N_y}
\int_{-\infty}^{0}
\!\!\!\!
 d\omega 
 \:
  {\rm Tr}_{\sigma} 
\Big[
 2t  
\sin{k_y} 
\sigma_0
\langle c_{n,k_y,\sigma}c^{\dag}_{n,k_y,\sigma} \rangle     
\\
&
\hspace{-0.15cm}
-{\rm g}
\cos{k_y} 
\Big(
\alpha 
 \sigma_x 
 \langle c_{n,k_y,\sigma}c^{\dag}_{n,k_y,\sigma} \rangle 
 -\beta
  \sigma_y
  \langle c_{n,k_y,\sigma}c^{\dag}_{n,k_y,\sigma} \rangle
  \Big)
  \Big].
\end{aligned}
\label{Eq:Topo_Current_1}
\end{align}
%
%
Using the definition of the momentum- and spin-resolved spectral functions, Eqs.~(\ref{Eq:Momentum_Resolved_LDOS}~and~\ref{Eq:Spin_Resolved_LDOS}), and  Matsubara Green's function, 
 $G_{nn}(k_y,{\mi} \omega)=\langle c^{}_{n,k_y,\sigma} c^{\dag}_{n,k_y,\sigma} \rangle,$
  the topological current along $y$-direction at the interface is given by
%
\begin{align}
\begin{aligned}
I_y
&
\!=\!
\sum_{n=1}^{N_x/2}
\sum_{k_y}
\frac{e}{N_y}
\int_{-\infty}^{0} d\omega 
\Big[ 
2t 
\sin{k_y}
\;
\rho_n(k_y,\omega)
\\
&
\hspace{2.05cm}
-{\rm g}
 \cos{k_y}
  \Big(
  \alpha  \rho_n^x(k_y,\omega)-\beta   \rho_n^y(k_y,\omega)
  \Big) 
 \Big].
\end{aligned}
\label{Eq:Topo_Current}
\end{align}
%
%
The first term in the above equation %
 is the contribution of the nearest-neighbor hopping in tight-binding model, whereas the second and third terms originate from Rashba and \DsH~SOCs, respectively.
Fig.~\ref{fig:Fig6} %
displays the topological 
 nonchiral
  charge current at the interface of the NCS-FM junction for the triplet-dominant $s+p$ (a), the triplet-dominant $d_{xy}+p$ (b) and the singlet-dominant $d_{xy}+p$ (c). 
 We do not show  the singlet-dominant $s+p$-case  because of its trivial topology.
In general, although for both cases of pure Rashba and mixed Rashba-\DsH~the 
  charge current increases with  field strength, but it shows negligible dependency for pure \DsH. 
 The magnitude of the 
   charge current for pure Rashba has maximum variation particularly for the triplet-dominant $d_{xy}+p$, but   by introducing the \DsH, it starts to decrease, and ends up to zero for pure \DsH. 
 Moreover, switching  the exchange field reverses  the direction of 
  the nonchiral current.\\

\begin{figure}[t]
 \begin{center}
\vspace{0.cm}
 \includegraphics[width=0.4 \linewidth]{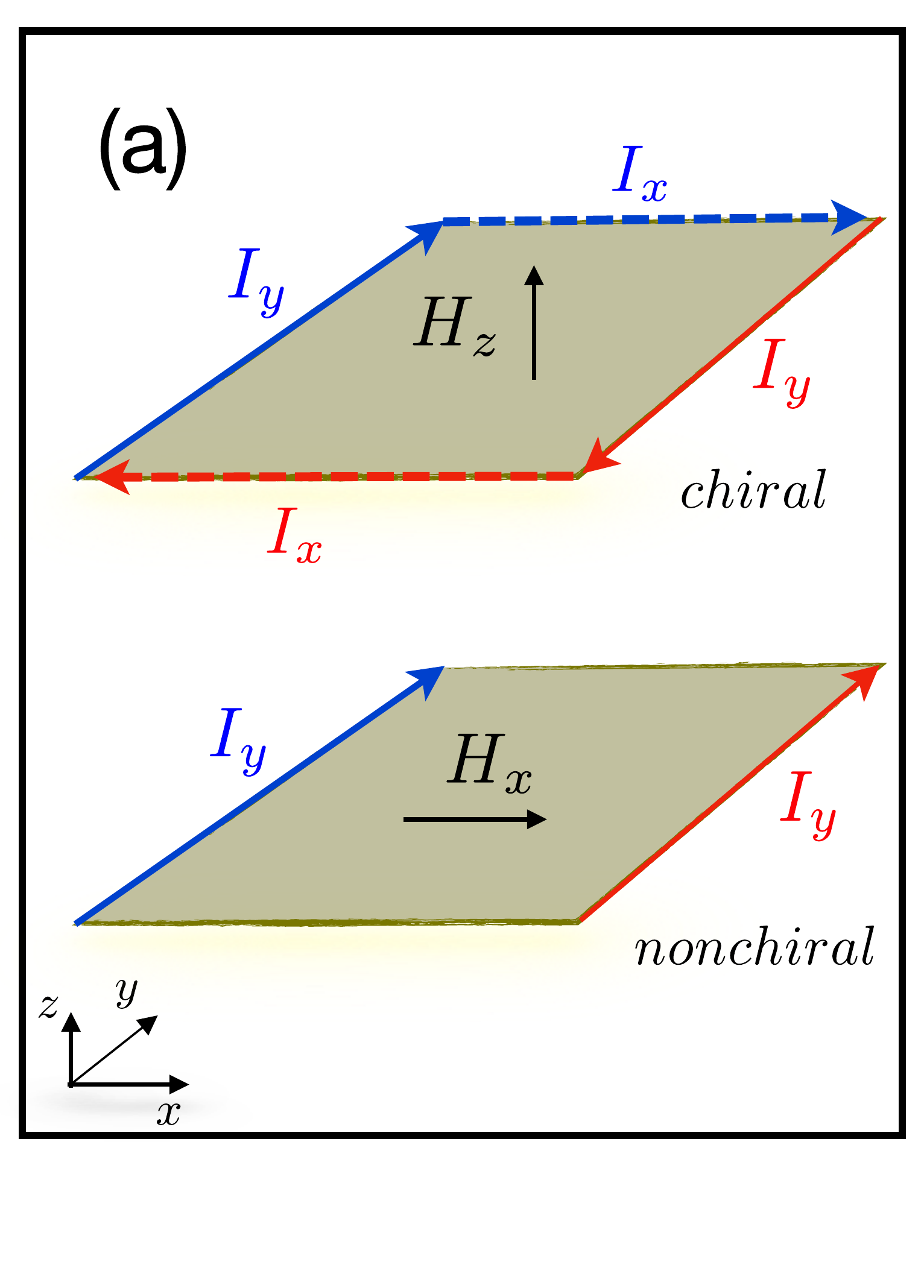} 
  \hspace{-.05cm}
  \includegraphics[width=0.58\linewidth]{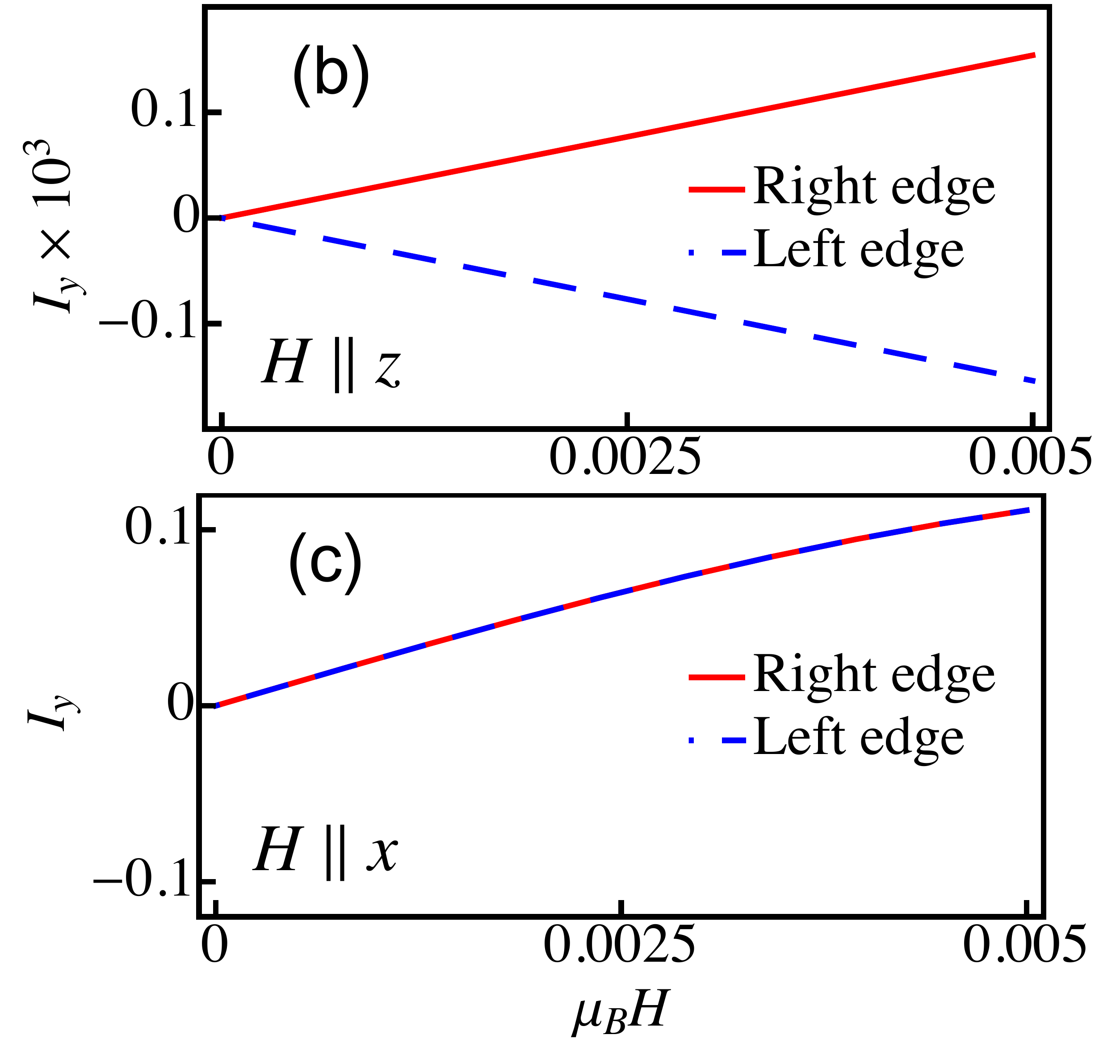} 
\end{center}
 \vspace{-0.7cm}
\caption{(Color online) 
(a) Representation of the topological current at the right (red) and left (blue) edges of a noncentrosymmetric superconductor with pure Rashba spin-orbit coupling and an exchange field, which is aligned  
in $z$-direction for  the upper panel, and in the  $x$-direction for the lower panel. For the cases of $H_z$ and $H_x$ the topological currents are chiral and nonchiral (unidirectional), respectively. 
(b) and (c) display topological currents at interface of NCS-FM junction for triplet-dominant  ($r=0.25$) with $d_{xy}+p$-wave superconducting states for 
chiral and nonchiral situations, respectively. 
 The currents have the same direction  for the field aligned in $x$-direction, $H_x$.
The current unit is given by $et/(2\hbar )$.
}
\label{fig:Fig7}
\end{figure}

Finally these results for the edge currents may be supported by symmetry arguments. In the BdG Hamiltionian without field the charge conjugation ${\cal C}= \tau_x \sigma_0  {\cal K}$ and time reversal are ${\cal T}={\mi} \tau_0 \sigma_y {\cal K} $  are preserved symmetries \cite{Matsuura:2013}  where ${\cal K}$ represents the complex conjugation operator. Their product, the chirality 
$\chi={\cal CT}={\mi}\tau_x\sigma_y $ 
is then also a  symmetry. It expresses the equivalence of   
positive and negative energy states in the BdG Hamiltonian according to 
$\chi{\cal H(\bk)} \chi^{-1}=-{\cal H(\bk)}$. 
The Zeeman part in Eq.~(\ref{Eq:Bulk_Ham}) which preserves  ${\cal C}$,  however, breaks  ${\cal T}$ and hence $\chi$ because 
$\chi{\cal H}_Z \chi^{-1}={\cal H}_Z$.
This symmetry breaking  leads to the apperance of the edge currents calculated explicitly before. They have a different pattern for in-plane and out-of-plane fields. A schematic sketch and a comparison of these two cases is shown in Fig.~\ref{fig:Fig7}.
In particular, in the Fig.~\ref{fig:Fig7}(a)  the edge current pattern for both situations is indicated, showing chiral and nonchiral (unidirectional)   currents, respectively. This leads to a field dependent current indicated in the Fig.~\ref{fig:Fig7}(b\&c).

%

\section{Conclusion}
We study the topological phases of a noncentrosymmetric superconductor in the presence of Rashba and/or \DsH~SOCs, which are characterized by gapless edge states. 
 Determination of the  Fermi surface for different strength of Rashba or \DsH~ shows that the helical split-bands touch each other at [$1\bar{1}$]- or [$11$]-directions for the condition  of  $|\alpha|=|\beta|$.
We investigate the structure function of dispersive and flat band edge states inside the superconducting gap, indicative of a generally non-trivial topology of the SC state.
 However, in the special case  $|\alpha|=|\beta|$ for an $s+p$-wave NCS  we obtain trivial topology without edge states,  even for a triplet-dominant state. \\

In addition we analyze in detail their spin texture for the cases of $s+p$ and $d_{xy}+p$ mixed parity gap functions with time reversal symmetry.
 We observe two counter-propagating modes flowing at each edge with opposite spin polarization.
  We also investigate the effect of a homogeneous exchange field on the edge states and find that it introduces 
an asymmetry in left- and right propagating edge states.\\

As a result of this coupling to an external exchange field along $x$-direction to the spin polarization of the topological edge modes, an energy gradient,  i.e. a finite group velocity  occurs for the originally flat band modes along $k_y$.
 This leads to emergence of 
 a nonchiral
 charge current
  unidirectional along both edges in
   $y$-direction carried by the surface states which changes its direction with that of the exchange field.
In this context we observe that 
 a nonchiral
  current is generated whenever  the NCS has non-zero spin polarization component along the exchange field.
Calculation of 
charge
 current in the presence of an exchange field along $x$-direction shows that its magnitude is maximum in a triplet-dominant $d_{xy}+p$-wave NCS with pure Rashba SOC. \\[1cm]

\section*{ACKNOWLEDGMENTS}
 We are grateful to P.~Fulde, A.~P.~Schnyder, K.\;S.~Kim and A.~Leonhardt  for fruitful discussions.
 This work  is supported through NRF funded by
 MSIP of Korea (2015R1C1A1A01052411) and (2017R1D1A1B03033465).
M.\ B.  is partially supported by the Abdus Salam International Centre for Theoretical Physics (ICTP, Trieste, Italy)  through the fund No. AF-03/18-01.
A.\ A. acknowledges the Max Planck POSTECH / KOREA Research Initiative (No. 2011-0031558)
 programs through NRF funded by MSIP of Korea.

\bibliography{References}

\begin{thebibliography}{83}%
\makeatletter
\providecommand \@ifxundefined [1]{%
 \@ifx{#1\undefined}
}%
\providecommand \@ifnum [1]{%
 \ifnum #1\expandafter \@firstoftwo
 \else \expandafter \@secondoftwo
 \fi
}%
\providecommand \@ifx [1]{%
 \ifx #1\expandafter \@firstoftwo
 \else \expandafter \@secondoftwo
 \fi
}%
\providecommand \natexlab [1]{#1}%
\providecommand \enquote  [1]{``#1''}%
\providecommand \bibnamefont  [1]{#1}%
\providecommand \bibfnamefont [1]{#1}%
\providecommand \citenamefont [1]{#1}%
\providecommand \href@noop [0]{\@secondoftwo}%
\providecommand \href [0]{\begingroup \@sanitize@url \@href}%
\providecommand \@href[1]{\@@startlink{#1}\@@href}%
\providecommand \@@href[1]{\endgroup#1\@@endlink}%
\providecommand \@sanitize@url [0]{\catcode `\\12\catcode `\$12\catcode
  `\&12\catcode `\#12\catcode `\^12\catcode `\_12\catcode `\%12\relax}%
\providecommand \@@startlink[1]{}%
\providecommand \@@endlink[0]{}%
\providecommand \url  [0]{\begingroup\@sanitize@url \@url }%
\providecommand \@url [1]{\endgroup\@href {#1}{\urlprefix }}%
\providecommand \urlprefix  [0]{URL }%
\providecommand \Eprint [0]{\href }%
\providecommand \doibase [0]{http://dx.doi.org/}%
\providecommand \selectlanguage [0]{\@gobble}%
\providecommand \bibinfo  [0]{\@secondoftwo}%
\providecommand \bibfield  [0]{\@secondoftwo}%
\providecommand \translation [1]{[#1]}%
\providecommand \BibitemOpen [0]{}%
\providecommand \bibitemStop [0]{}%
\providecommand \bibitemNoStop [0]{.\EOS\space}%
\providecommand \EOS [0]{\spacefactor3000\relax}%
\providecommand \BibitemShut  [1]{\csname bibitem#1\endcsname}%
\let\auto@bib@innerbib\@empty
\bibitem [{\citenamefont {Hasan}\ and\ \citenamefont
  {Kane}(2010)}]{Zahid:2010}%
  \BibitemOpen
  \bibfield  {author} {\bibinfo {author} {\bibfnamefont {M.~Z.}\ \bibnamefont
  {Hasan}}\ and\ \bibinfo {author} {\bibfnamefont {C.~L.}\ \bibnamefont
  {Kane}},\ }\href {\doibase 10.1103/RevModPhys.82.3045} {\bibfield  {journal}
  {\bibinfo  {journal} {Rev. Mod. Phys.}\ }\textbf {\bibinfo {volume} {82}},\
  \bibinfo {pages} {3045} (\bibinfo {year} {2010})}\BibitemShut {NoStop}%
\bibitem [{\citenamefont {Qi}\ and\ \citenamefont {Zhang}(2011)}]{Qi:2011}%
  \BibitemOpen
  \bibfield  {author} {\bibinfo {author} {\bibfnamefont {X.-L.}\ \bibnamefont
  {Qi}}\ and\ \bibinfo {author} {\bibfnamefont {S.-C.}\ \bibnamefont {Zhang}},\
  }\href {\doibase 10.1103/RevModPhys.83.1057} {\bibfield  {journal} {\bibinfo
  {journal} {Rev. Mod. Phys.}\ }\textbf {\bibinfo {volume} {83}},\ \bibinfo
  {pages} {1057} (\bibinfo {year} {2011})}\BibitemShut {NoStop}%
\bibitem [{\citenamefont {Sato}\ and\ \citenamefont {Ando}(2017)}]{Sato:2017}%
  \BibitemOpen
  \bibfield  {author} {\bibinfo {author} {\bibfnamefont {M.}~\bibnamefont
  {Sato}}\ and\ \bibinfo {author} {\bibfnamefont {Y.}~\bibnamefont {Ando}},\
  }\href {http://stacks.iop.org/0034-4885/80/i=7/a=076501} {\bibfield
  {journal} {\bibinfo  {journal} {Reports on Progress in Physics}\ }\textbf
  {\bibinfo {volume} {80}},\ \bibinfo {pages} {76501} (\bibinfo {year}
  {2017})}\BibitemShut {NoStop}%
\bibitem [{\citenamefont {Zare}\ \emph {et~al.}(2017)\citenamefont {Zare},
  \citenamefont {Biderang},\ and\ \citenamefont {Akbari}}]{Zare:2017}%
  \BibitemOpen
  \bibfield  {author} {\bibinfo {author} {\bibfnamefont {M.-H.}\ \bibnamefont
  {Zare}}, \bibinfo {author} {\bibfnamefont {M.}~\bibnamefont {Biderang}}, \
  and\ \bibinfo {author} {\bibfnamefont {A.}~\bibnamefont {Akbari}},\ }\href
  {\doibase 10.1103/PhysRevB.96.205156} {\bibfield  {journal} {\bibinfo
  {journal} {Phys. Rev. B}\ }\textbf {\bibinfo {volume} {96}},\ \bibinfo
  {pages} {205156} (\bibinfo {year} {2017})}\BibitemShut {NoStop}%
\bibitem [{\citenamefont {Lambert}\ \emph {et~al.}(2017)\citenamefont
  {Lambert}, \citenamefont {Akbari.}, \citenamefont {Thalmeier},\ and\
  \citenamefont {Eremin}}]{Lambert:2017}%
  \BibitemOpen
  \bibfield  {author} {\bibinfo {author} {\bibfnamefont {F.}~\bibnamefont
  {Lambert}}, \bibinfo {author} {\bibfnamefont {A.}~\bibnamefont {Akbari.}},
  \bibinfo {author} {\bibfnamefont {P.}~\bibnamefont {Thalmeier}}, \ and\
  \bibinfo {author} {\bibfnamefont {I.}~\bibnamefont {Eremin}},\ }\href
  {\doibase 10.1103/PhysRevLett.118.087004} {\bibfield  {journal} {\bibinfo
  {journal} {Phys. Rev. Lett.}\ }\textbf {\bibinfo {volume} {118}},\ \bibinfo
  {pages} {087004} (\bibinfo {year} {2017})}\BibitemShut {NoStop}%
\bibitem [{\citenamefont {Wan}\ \emph {et~al.}(2011)\citenamefont {Wan},
  \citenamefont {Turner}, \citenamefont {Vishwanath},\ and\ \citenamefont
  {Savrasov}}]{Wan:2011}%
  \BibitemOpen
  \bibfield  {author} {\bibinfo {author} {\bibfnamefont {X.}~\bibnamefont
  {Wan}}, \bibinfo {author} {\bibfnamefont {A.~M.}\ \bibnamefont {Turner}},
  \bibinfo {author} {\bibfnamefont {A.}~\bibnamefont {Vishwanath}}, \ and\
  \bibinfo {author} {\bibfnamefont {S.~Y.}\ \bibnamefont {Savrasov}},\ }\href
  {\doibase 10.1103/PhysRevB.83.205101} {\bibfield  {journal} {\bibinfo
  {journal} {Phys. Rev. B}\ }\textbf {\bibinfo {volume} {83}},\ \bibinfo
  {pages} {205101} (\bibinfo {year} {2011})}\BibitemShut {NoStop}%
\bibitem [{\citenamefont {Turner}\ and\ \citenamefont
  {Vishwanath}(2013)}]{Turner:2013}%
  \BibitemOpen
  \bibfield  {author} {\bibinfo {author} {\bibfnamefont {A.}~\bibnamefont
  {Turner}}\ and\ \bibinfo {author} {\bibfnamefont {A.}~\bibnamefont
  {Vishwanath}},\ }\href {https://books.google.com/books?id=1K0ZDAAAQBAJ}
  {\emph {\bibinfo {title} {{Topological Insulators: Chapter 11. Beyond Band
  Insulators: Topology of Semimetals and Interacting Phases}}}},\ {Contemporary
  Concepts of Condensed Matter Science}\ (\bibinfo  {publisher} {Elsevier
  Science},\ \bibinfo {year} {2013})\BibitemShut {NoStop}%
\bibitem [{\citenamefont {Yan}\ and\ \citenamefont {Felser}(2017)}]{Yan:2017}%
  \BibitemOpen
  \bibfield  {author} {\bibinfo {author} {\bibfnamefont {B.}~\bibnamefont
  {Yan}}\ and\ \bibinfo {author} {\bibfnamefont {C.}~\bibnamefont {Felser}},\
  }\href {\doibase 10.1146/annurev-conmatphys-031016-025458} {\bibfield
  {journal} {\bibinfo  {journal} {Annual Review of Condensed Matter Physics}\
  }\textbf {\bibinfo {volume} {8}},\ \bibinfo {pages} {337} (\bibinfo {year}
  {2017})}\BibitemShut {NoStop}%
\bibitem [{\citenamefont {Hasan}\ \emph {et~al.}(2017)\citenamefont {Hasan},
  \citenamefont {Xu}, \citenamefont {Belopolski},\ and\ \citenamefont
  {Huang}}]{Zahid:2017}%
  \BibitemOpen
  \bibfield  {author} {\bibinfo {author} {\bibfnamefont {M.~Z.}\ \bibnamefont
  {Hasan}}, \bibinfo {author} {\bibfnamefont {S.-Y.}\ \bibnamefont {Xu}},
  \bibinfo {author} {\bibfnamefont {I.}~\bibnamefont {Belopolski}}, \ and\
  \bibinfo {author} {\bibfnamefont {S.-M.}\ \bibnamefont {Huang}},\ }\href
  {\doibase 10.1146/annurev-conmatphys-031016-025225} {\bibfield  {journal}
  {\bibinfo  {journal} {Annual Review of Condensed Matter Physics}\ }\textbf
  {\bibinfo {volume} {8}},\ \bibinfo {pages} {289} (\bibinfo {year}
  {2017})}\BibitemShut {NoStop}%
\bibitem [{\citenamefont {{Biderang}}\ \emph {et~al.}(2018)\citenamefont
  {{Biderang}}, \citenamefont {{Leonhardt}}, \citenamefont {{Raghuvanshi}},
  \citenamefont {{Schnyder}},\ and\ \citenamefont
  {{Akbari}}}]{Biderang:2018aa}%
  \BibitemOpen
  \bibfield  {author} {\bibinfo {author} {\bibfnamefont {M.}~\bibnamefont
  {{Biderang}}}, \bibinfo {author} {\bibfnamefont {A.}~\bibnamefont
  {{Leonhardt}}}, \bibinfo {author} {\bibfnamefont {N.}~\bibnamefont
  {{Raghuvanshi}}}, \bibinfo {author} {\bibfnamefont {A.~P.}\ \bibnamefont
  {{Schnyder}}}, \ and\ \bibinfo {author} {\bibfnamefont {A.}~\bibnamefont
  {{Akbari}}},\ }\href@noop {} {\bibfield  {journal} {\bibinfo  {journal}
  {ArXiv e-prints}\ } (\bibinfo {year} {2018})},\ \Eprint
  {http://arxiv.org/abs/1802.09139} {arXiv:1802.09139 [cond-mat.str-el]}
  \BibitemShut {NoStop}%
\bibitem [{\citenamefont {Nayak}\ \emph {et~al.}(2008)\citenamefont {Nayak},
  \citenamefont {Simon}, \citenamefont {Stern}, \citenamefont {Freedman},\ and\
  \citenamefont {{Das Sarma}}}]{Nayak:2008}%
  \BibitemOpen
  \bibfield  {author} {\bibinfo {author} {\bibfnamefont {C.}~\bibnamefont
  {Nayak}}, \bibinfo {author} {\bibfnamefont {S.~H.}\ \bibnamefont {Simon}},
  \bibinfo {author} {\bibfnamefont {A.}~\bibnamefont {Stern}}, \bibinfo
  {author} {\bibfnamefont {M.}~\bibnamefont {Freedman}}, \ and\ \bibinfo
  {author} {\bibfnamefont {S.}~\bibnamefont {{Das Sarma}}},\ }\href@noop {}
  {\bibfield  {journal} {\bibinfo  {journal} {Rev. Mod. Phys.}\ }\textbf
  {\bibinfo {volume} {80}},\ \bibinfo {pages} {1083} (\bibinfo {year}
  {2008})}\BibitemShut {NoStop}%
\bibitem [{\citenamefont {Terhal}(2015)}]{Terhal:2015}%
  \BibitemOpen
  \bibfield  {author} {\bibinfo {author} {\bibfnamefont {B.~M.}\ \bibnamefont
  {Terhal}},\ }\href {\doibase 10.1103/RevModPhys.87.307} {\bibfield  {journal}
  {\bibinfo  {journal} {Rev. Mod. Phys.}\ }\textbf {\bibinfo {volume} {87}},\
  \bibinfo {pages} {307} (\bibinfo {year} {2015})}\BibitemShut {NoStop}%
\bibitem [{\citenamefont {Alicea}(2010)}]{Alicea:2010}%
  \BibitemOpen
  \bibfield  {author} {\bibinfo {author} {\bibfnamefont {J.}~\bibnamefont
  {Alicea}},\ }\href {\doibase 10.1103/PhysRevB.81.125318} {\bibfield
  {journal} {\bibinfo  {journal} {Phys. Rev. B}\ }\textbf {\bibinfo {volume}
  {81}},\ \bibinfo {pages} {125318} (\bibinfo {year} {2010})}\BibitemShut
  {NoStop}%
\bibitem [{\citenamefont {Bauer}\ and\ \citenamefont
  {Sigrist}(2012)}]{Bauer:2012}%
  \BibitemOpen
  \bibfield  {author} {\bibinfo {author} {\bibfnamefont {E.}~\bibnamefont
  {Bauer}}\ and\ \bibinfo {author} {\bibfnamefont {M.}~\bibnamefont
  {Sigrist}},\ }\href {https://books.google.com/books?id=nDZ4lKD00t8C} {\emph
  {\bibinfo {title} {{Non-Centrosymmetric Superconductors: Introduction and
  Overview}}}},\ {Lecture Notes in Physics}\ (\bibinfo  {publisher} {Springer
  Berlin Heidelberg},\ \bibinfo {year} {2012})\BibitemShut {NoStop}%
\bibitem [{\citenamefont {Schnyder}\ and\ \citenamefont
  {Brydon}(2015)}]{Schnyder:2015}%
  \BibitemOpen
  \bibfield  {author} {\bibinfo {author} {\bibfnamefont {A.~P.}\ \bibnamefont
  {Schnyder}}\ and\ \bibinfo {author} {\bibfnamefont {P.~M.~R.}\ \bibnamefont
  {Brydon}},\ }\href {http://stacks.iop.org/0953-8984/27/i=24/a=243201}
  {\bibfield  {journal} {\bibinfo  {journal} {Journal of Physics: Condensed
  Matter}\ }\textbf {\bibinfo {volume} {27}},\ \bibinfo {pages} {243201}
  (\bibinfo {year} {2015})}\BibitemShut {NoStop}%
\bibitem [{\citenamefont {Sato}(2006)}]{Sato:2006}%
  \BibitemOpen
  \bibfield  {author} {\bibinfo {author} {\bibfnamefont {M.}~\bibnamefont
  {Sato}},\ }\href {\doibase 10.1103/PhysRevB.73.214502} {\bibfield  {journal}
  {\bibinfo  {journal} {Phys. Rev. B}\ }\textbf {\bibinfo {volume} {73}},\
  \bibinfo {pages} {214502} (\bibinfo {year} {2006})}\BibitemShut {NoStop}%
\bibitem [{\citenamefont {Qi}\ \emph {et~al.}(2010)\citenamefont {Qi},
  \citenamefont {Hughes},\ and\ \citenamefont {Zhang}}]{Qi:2010}%
  \BibitemOpen
  \bibfield  {author} {\bibinfo {author} {\bibfnamefont {X.-L.}\ \bibnamefont
  {Qi}}, \bibinfo {author} {\bibfnamefont {T.~L.}\ \bibnamefont {Hughes}}, \
  and\ \bibinfo {author} {\bibfnamefont {S.-C.}\ \bibnamefont {Zhang}},\ }\href
  {\doibase 10.1103/PhysRevB.81.134508} {\bibfield  {journal} {\bibinfo
  {journal} {Phys. Rev. B}\ }\textbf {\bibinfo {volume} {81}},\ \bibinfo
  {pages} {134508} (\bibinfo {year} {2010})}\BibitemShut {NoStop}%
\bibitem [{\citenamefont {Sato}\ \emph {et~al.}(2011)\citenamefont {Sato},
  \citenamefont {Tanaka}, \citenamefont {Yada},\ and\ \citenamefont
  {Yokoyama}}]{Sato:2011}%
  \BibitemOpen
  \bibfield  {author} {\bibinfo {author} {\bibfnamefont {M.}~\bibnamefont
  {Sato}}, \bibinfo {author} {\bibfnamefont {Y.}~\bibnamefont {Tanaka}},
  \bibinfo {author} {\bibfnamefont {K.}~\bibnamefont {Yada}}, \ and\ \bibinfo
  {author} {\bibfnamefont {T.}~\bibnamefont {Yokoyama}},\ }\href {\doibase
  10.1103/PhysRevB.83.224511} {\bibfield  {journal} {\bibinfo  {journal} {Phys.
  Rev. B}\ }\textbf {\bibinfo {volume} {83}},\ \bibinfo {pages} {224511}
  (\bibinfo {year} {2011})}\BibitemShut {NoStop}%
\bibitem [{\citenamefont {Schnyder}\ and\ \citenamefont
  {Ryu}(2011)}]{Schnyder:2011}%
  \BibitemOpen
  \bibfield  {author} {\bibinfo {author} {\bibfnamefont {A.~P.}\ \bibnamefont
  {Schnyder}}\ and\ \bibinfo {author} {\bibfnamefont {S.}~\bibnamefont {Ryu}},\
  }\href {\doibase 10.1103/PhysRevB.84.060504} {\bibfield  {journal} {\bibinfo
  {journal} {Phys. Rev. B}\ }\textbf {\bibinfo {volume} {84}},\ \bibinfo
  {pages} {060504} (\bibinfo {year} {2011})}\BibitemShut {NoStop}%
\bibitem [{\citenamefont {Yip}(2014)}]{Yip:2014}%
  \BibitemOpen
  \bibfield  {author} {\bibinfo {author} {\bibfnamefont {S.}~\bibnamefont
  {Yip}},\ }\href {\doibase 10.1146/annurev-conmatphys-031113-133912}
  {\bibfield  {journal} {\bibinfo  {journal} {Annual Review of Condensed Matter
  Physics}\ }\textbf {\bibinfo {volume} {5}},\ \bibinfo {pages} {15} (\bibinfo
  {year} {2014})}\BibitemShut {NoStop}%
\bibitem [{\citenamefont {Yada}\ \emph {et~al.}(2011)\citenamefont {Yada},
  \citenamefont {Sato}, \citenamefont {Tanaka},\ and\ \citenamefont
  {Yokoyama}}]{Yada:2011}%
  \BibitemOpen
  \bibfield  {author} {\bibinfo {author} {\bibfnamefont {K.}~\bibnamefont
  {Yada}}, \bibinfo {author} {\bibfnamefont {M.}~\bibnamefont {Sato}}, \bibinfo
  {author} {\bibfnamefont {Y.}~\bibnamefont {Tanaka}}, \ and\ \bibinfo {author}
  {\bibfnamefont {T.}~\bibnamefont {Yokoyama}},\ }\href {\doibase
  10.1103/PhysRevB.83.064505} {\bibfield  {journal} {\bibinfo  {journal} {Phys.
  Rev. B}\ }\textbf {\bibinfo {volume} {83}},\ \bibinfo {pages} {064505}
  (\bibinfo {year} {2011})}\BibitemShut {NoStop}%
\bibitem [{\citenamefont {Schnyder}\ \emph {et~al.}(2012)\citenamefont
  {Schnyder}, \citenamefont {Brydon},\ and\ \citenamefont
  {Timm}}]{Schnyder:2012}%
  \BibitemOpen
  \bibfield  {author} {\bibinfo {author} {\bibfnamefont {A.~P.}\ \bibnamefont
  {Schnyder}}, \bibinfo {author} {\bibfnamefont {P.~M.~R.}\ \bibnamefont
  {Brydon}}, \ and\ \bibinfo {author} {\bibfnamefont {C.}~\bibnamefont
  {Timm}},\ }\href {\doibase 10.1103/PhysRevB.85.024522} {\bibfield  {journal}
  {\bibinfo  {journal} {Phys. Rev. B}\ }\textbf {\bibinfo {volume} {85}},\
  \bibinfo {pages} {024522} (\bibinfo {year} {2012})}\BibitemShut {NoStop}%
\bibitem [{\citenamefont {Dahlhaus}\ \emph {et~al.}(2012)\citenamefont
  {Dahlhaus}, \citenamefont {Gibertini},\ and\ \citenamefont
  {Beenakker}}]{Dahlhaus:2012}%
  \BibitemOpen
  \bibfield  {author} {\bibinfo {author} {\bibfnamefont {J.~P.}\ \bibnamefont
  {Dahlhaus}}, \bibinfo {author} {\bibfnamefont {M.}~\bibnamefont {Gibertini}},
  \ and\ \bibinfo {author} {\bibfnamefont {C.~W.~J.}\ \bibnamefont
  {Beenakker}},\ }\href {\doibase 10.1103/PhysRevB.86.174520} {\bibfield
  {journal} {\bibinfo  {journal} {Phys. Rev. B}\ }\textbf {\bibinfo {volume}
  {86}},\ \bibinfo {pages} {174520} (\bibinfo {year} {2012})}\BibitemShut
  {NoStop}%
\bibitem [{\citenamefont {Matsuura}\ \emph {et~al.}(2013)\citenamefont
  {Matsuura}, \citenamefont {Chang}, \citenamefont {Schnyder},\ and\
  \citenamefont {Ryu}}]{Matsuura:2013}%
  \BibitemOpen
  \bibfield  {author} {\bibinfo {author} {\bibfnamefont {S.}~\bibnamefont
  {Matsuura}}, \bibinfo {author} {\bibfnamefont {P.-Y.}\ \bibnamefont {Chang}},
  \bibinfo {author} {\bibfnamefont {A.~P.}\ \bibnamefont {Schnyder}}, \ and\
  \bibinfo {author} {\bibfnamefont {S.}~\bibnamefont {Ryu}},\ }\href
  {http://stacks.iop.org/1367-2630/15/i=6/a=065001} {\bibfield  {journal}
  {\bibinfo  {journal} {New Journal of Physics}\ }\textbf {\bibinfo {volume}
  {15}},\ \bibinfo {pages} {065001} (\bibinfo {year} {2013})}\BibitemShut
  {NoStop}%
\bibitem [{\citenamefont {Daido}\ and\ \citenamefont
  {Yanase}(2016)}]{Daido:2016aa}%
  \BibitemOpen
  \bibfield  {author} {\bibinfo {author} {\bibfnamefont {A.}~\bibnamefont
  {Daido}}\ and\ \bibinfo {author} {\bibfnamefont {Y.}~\bibnamefont {Yanase}},\
  }\href {\doibase 10.1103/PhysRevB.94.054519} {\bibfield  {journal} {\bibinfo
  {journal} {Phys. Rev. B}\ }\textbf {\bibinfo {volume} {94}},\ \bibinfo
  {pages} {054519} (\bibinfo {year} {2016})}\BibitemShut {NoStop}%
\bibitem [{\citenamefont {Daido}\ and\ \citenamefont
  {Yanase}(2017)}]{Daido:2017aa}%
  \BibitemOpen
  \bibfield  {author} {\bibinfo {author} {\bibfnamefont {A.}~\bibnamefont
  {Daido}}\ and\ \bibinfo {author} {\bibfnamefont {Y.}~\bibnamefont {Yanase}},\
  }\href {\doibase 10.1103/PhysRevB.95.134507} {\bibfield  {journal} {\bibinfo
  {journal} {Phys. Rev. B}\ }\textbf {\bibinfo {volume} {95}},\ \bibinfo
  {pages} {134507} (\bibinfo {year} {2017})}\BibitemShut {NoStop}%
\bibitem [{\citenamefont {Timm}\ \emph {et~al.}(2015)\citenamefont {Timm},
  \citenamefont {Rex},\ and\ \citenamefont {Brydon}}]{Timm:2015}%
  \BibitemOpen
  \bibfield  {author} {\bibinfo {author} {\bibfnamefont {C.}~\bibnamefont
  {Timm}}, \bibinfo {author} {\bibfnamefont {S.}~\bibnamefont {Rex}}, \ and\
  \bibinfo {author} {\bibfnamefont {P.~M.~R.}\ \bibnamefont {Brydon}},\ }\href
  {\doibase 10.1103/PhysRevB.91.180503} {\bibfield  {journal} {\bibinfo
  {journal} {Phys. Rev. B}\ }\textbf {\bibinfo {volume} {91}},\ \bibinfo
  {pages} {180503} (\bibinfo {year} {2015})}\BibitemShut {NoStop}%
\bibitem [{\citenamefont {Tanaka}\ \emph {et~al.}(2012)\citenamefont {Tanaka},
  \citenamefont {Sato},\ and\ \citenamefont {Nagaosa}}]{Tanaka:2012}%
  \BibitemOpen
  \bibfield  {author} {\bibinfo {author} {\bibfnamefont {Y.}~\bibnamefont
  {Tanaka}}, \bibinfo {author} {\bibfnamefont {M.}~\bibnamefont {Sato}}, \ and\
  \bibinfo {author} {\bibfnamefont {N.}~\bibnamefont {Nagaosa}},\ }\href
  {\doibase 10.1143/JPSJ.81.011013} {\bibfield  {journal} {\bibinfo  {journal}
  {Journal of the Physical Society of Japan}\ }\textbf {\bibinfo {volume}
  {81}},\ \bibinfo {pages} {011013} (\bibinfo {year} {2012})}\BibitemShut
  {NoStop}%
\bibitem [{\citenamefont {Tang}\ and\ \citenamefont {Fu}(2014)}]{Tang:2014}%
  \BibitemOpen
  \bibfield  {author} {\bibinfo {author} {\bibfnamefont {E.}~\bibnamefont
  {Tang}}\ and\ \bibinfo {author} {\bibfnamefont {L.}~\bibnamefont {Fu}},\
  }\href {\doibase 10.1038/NPHYS3109} {\  (\bibinfo {year} {2014}),\
  10.1038/NPHYS3109}\BibitemShut {NoStop}%
\bibitem [{\citenamefont {Chiu}\ and\ \citenamefont
  {Schnyder}(2014)}]{Kai:2014}%
  \BibitemOpen
  \bibfield  {author} {\bibinfo {author} {\bibfnamefont {C.-K.}\ \bibnamefont
  {Chiu}}\ and\ \bibinfo {author} {\bibfnamefont {A.~P.}\ \bibnamefont
  {Schnyder}},\ }\href {\doibase 10.1103/PhysRevB.90.205136} {\bibfield
  {journal} {\bibinfo  {journal} {Phys. Rev. B}\ }\textbf {\bibinfo {volume}
  {90}},\ \bibinfo {pages} {205136} (\bibinfo {year} {2014})}\BibitemShut
  {NoStop}%
\bibitem [{\citenamefont {Beenakker}(2013)}]{Beenakker:2013}%
  \BibitemOpen
  \bibfield  {author} {\bibinfo {author} {\bibfnamefont {C.}~\bibnamefont
  {Beenakker}},\ }\href {\doibase 10.1146/annurev-conmatphys-030212-184337}
  {\bibfield  {journal} {\bibinfo  {journal} {Annual Review of Condensed Matter
  Physics}\ }\textbf {\bibinfo {volume} {4}},\ \bibinfo {pages} {113} (\bibinfo
  {year} {2013})}\BibitemShut {NoStop}%
\bibitem [{\citenamefont {Beenakker}(2015)}]{Beenakker:2015}%
  \BibitemOpen
  \bibfield  {author} {\bibinfo {author} {\bibfnamefont {C.~W.~J.}\
  \bibnamefont {Beenakker}},\ }\href {\doibase 10.1103/RevModPhys.87.1037}
  {\bibfield  {journal} {\bibinfo  {journal} {Rev. Mod. Phys.}\ }\textbf
  {\bibinfo {volume} {87}},\ \bibinfo {pages} {1037} (\bibinfo {year}
  {2015})}\BibitemShut {NoStop}%
\bibitem [{\citenamefont {Sato}\ and\ \citenamefont
  {Fujimoto}(2009)}]{Sato_Fujimoto:2009}%
  \BibitemOpen
  \bibfield  {author} {\bibinfo {author} {\bibfnamefont {M.}~\bibnamefont
  {Sato}}\ and\ \bibinfo {author} {\bibfnamefont {S.}~\bibnamefont
  {Fujimoto}},\ }\href {\doibase 10.1103/PhysRevB.79.094504} {\bibfield
  {journal} {\bibinfo  {journal} {Phys. Rev. B}\ }\textbf {\bibinfo {volume}
  {79}},\ \bibinfo {pages} {094504} (\bibinfo {year} {2009})}\BibitemShut
  {NoStop}%
\bibitem [{\citenamefont {Sato}\ \emph {et~al.}(2010)\citenamefont {Sato},
  \citenamefont {Takahashi},\ and\ \citenamefont {Fujimoto}}]{Sato:2010}%
  \BibitemOpen
  \bibfield  {author} {\bibinfo {author} {\bibfnamefont {M.}~\bibnamefont
  {Sato}}, \bibinfo {author} {\bibfnamefont {Y.}~\bibnamefont {Takahashi}}, \
  and\ \bibinfo {author} {\bibfnamefont {S.}~\bibnamefont {Fujimoto}},\ }\href
  {\doibase 10.1103/PhysRevB.82.134521} {\bibfield  {journal} {\bibinfo
  {journal} {Phys. Rev. B}\ }\textbf {\bibinfo {volume} {82}},\ \bibinfo
  {pages} {134521} (\bibinfo {year} {2010})}\BibitemShut {NoStop}%
\bibitem [{\citenamefont {Wong}\ \emph {et~al.}(2013)\citenamefont {Wong},
  \citenamefont {Liu}, \citenamefont {Law},\ and\ \citenamefont
  {Lee}}]{Wong:2013}%
  \BibitemOpen
  \bibfield  {author} {\bibinfo {author} {\bibfnamefont {C.~L.~M.}\
  \bibnamefont {Wong}}, \bibinfo {author} {\bibfnamefont {J.}~\bibnamefont
  {Liu}}, \bibinfo {author} {\bibfnamefont {K.~T.}\ \bibnamefont {Law}}, \ and\
  \bibinfo {author} {\bibfnamefont {P.~A.}\ \bibnamefont {Lee}},\ }\href
  {\doibase 10.1103/PhysRevB.88.060504} {\bibfield  {journal} {\bibinfo
  {journal} {Phys. Rev. B}\ }\textbf {\bibinfo {volume} {88}},\ \bibinfo
  {pages} {060504} (\bibinfo {year} {2013})}\BibitemShut {NoStop}%
\bibitem [{\citenamefont {Read}\ and\ \citenamefont {Green}(2000)}]{Read:2000}%
  \BibitemOpen
  \bibfield  {author} {\bibinfo {author} {\bibfnamefont {N.}~\bibnamefont
  {Read}}\ and\ \bibinfo {author} {\bibfnamefont {D.}~\bibnamefont {Green}},\
  }\href {\doibase 10.1103/PhysRevB.61.10267} {\bibfield  {journal} {\bibinfo
  {journal} {Phys. Rev. B}\ }\textbf {\bibinfo {volume} {61}},\ \bibinfo
  {pages} {10267} (\bibinfo {year} {2000})}\BibitemShut {NoStop}%
\bibitem [{\citenamefont {Law}\ \emph {et~al.}(2009)\citenamefont {Law},
  \citenamefont {Lee},\ and\ \citenamefont {Ng}}]{Law:2009}%
  \BibitemOpen
  \bibfield  {author} {\bibinfo {author} {\bibfnamefont {K.~T.}\ \bibnamefont
  {Law}}, \bibinfo {author} {\bibfnamefont {P.~A.}\ \bibnamefont {Lee}}, \ and\
  \bibinfo {author} {\bibfnamefont {T.~K.}\ \bibnamefont {Ng}},\ }\href
  {\doibase 10.1103/PhysRevLett.103.237001} {\bibfield  {journal} {\bibinfo
  {journal} {Phys. Rev. Lett.}\ }\textbf {\bibinfo {volume} {103}},\ \bibinfo
  {pages} {237001} (\bibinfo {year} {2009})}\BibitemShut {NoStop}%
\bibitem [{\citenamefont {Brydon}\ \emph {et~al.}(2013)\citenamefont {Brydon},
  \citenamefont {Timm},\ and\ \citenamefont {Schnyder}}]{Brydon:2013aa}%
  \BibitemOpen
  \bibfield  {author} {\bibinfo {author} {\bibfnamefont {P.~M.~R.}\
  \bibnamefont {Brydon}}, \bibinfo {author} {\bibfnamefont {C.}~\bibnamefont
  {Timm}}, \ and\ \bibinfo {author} {\bibfnamefont {A.~P.}\ \bibnamefont
  {Schnyder}},\ }\href {http://stacks.iop.org/1367-2630/15/i=4/a=045019}
  {\bibfield  {journal} {\bibinfo  {journal} {New Journal of Physics}\ }\textbf
  {\bibinfo {volume} {15}},\ \bibinfo {pages} {045019} (\bibinfo {year}
  {2013})}\BibitemShut {NoStop}%
\bibitem [{\citenamefont {Schnyder}\ \emph {et~al.}(2013)\citenamefont
  {Schnyder}, \citenamefont {Timm},\ and\ \citenamefont
  {Brydon}}]{Schnyder:2013aa}%
  \BibitemOpen
  \bibfield  {author} {\bibinfo {author} {\bibfnamefont {A.~P.}\ \bibnamefont
  {Schnyder}}, \bibinfo {author} {\bibfnamefont {C.}~\bibnamefont {Timm}}, \
  and\ \bibinfo {author} {\bibfnamefont {P.~M.~R.}\ \bibnamefont {Brydon}},\
  }\href {\doibase 10.1103/PhysRevLett.111.077001} {\bibfield  {journal}
  {\bibinfo  {journal} {Phys. Rev. Lett.}\ }\textbf {\bibinfo {volume} {111}},\
  \bibinfo {pages} {077001} (\bibinfo {year} {2013})}\BibitemShut {NoStop}%
\bibitem [{\citenamefont {Tanaka}\ \emph {et~al.}(2009)\citenamefont {Tanaka},
  \citenamefont {Yokoyama}, \citenamefont {Balatsky},\ and\ \citenamefont
  {Nagaosa}}]{Tanaka:2009:PRB}%
  \BibitemOpen
  \bibfield  {author} {\bibinfo {author} {\bibfnamefont {Y.}~\bibnamefont
  {Tanaka}}, \bibinfo {author} {\bibfnamefont {T.}~\bibnamefont {Yokoyama}},
  \bibinfo {author} {\bibfnamefont {A.~V.}\ \bibnamefont {Balatsky}}, \ and\
  \bibinfo {author} {\bibfnamefont {N.}~\bibnamefont {Nagaosa}},\ }\href
  {\doibase 10.1103/PhysRevB.79.060505} {\bibfield  {journal} {\bibinfo
  {journal} {Phys. Rev. B}\ }\textbf {\bibinfo {volume} {79}},\ \bibinfo
  {pages} {060505} (\bibinfo {year} {2009})}\BibitemShut {NoStop}%
\bibitem [{\citenamefont {Vorontsov}\ \emph {et~al.}(2008)\citenamefont
  {Vorontsov}, \citenamefont {Vekhter},\ and\ \citenamefont
  {Eschrig}}]{Vorontsov:2008:PRL}%
  \BibitemOpen
  \bibfield  {author} {\bibinfo {author} {\bibfnamefont {A.~B.}\ \bibnamefont
  {Vorontsov}}, \bibinfo {author} {\bibfnamefont {I.}~\bibnamefont {Vekhter}},
  \ and\ \bibinfo {author} {\bibfnamefont {M.}~\bibnamefont {Eschrig}},\ }\href
  {\doibase 10.1103/PhysRevLett.101.127003} {\bibfield  {journal} {\bibinfo
  {journal} {Phys. Rev. Lett.}\ }\textbf {\bibinfo {volume} {101}},\ \bibinfo
  {pages} {127003} (\bibinfo {year} {2008})}\BibitemShut {NoStop}%
\bibitem [{\citenamefont {Togano}\ \emph {et~al.}(2004)\citenamefont {Togano},
  \citenamefont {Badica}, \citenamefont {Nakamori}, \citenamefont {Orimo},
  \citenamefont {Takeya},\ and\ \citenamefont {Hirata}}]{Togano:2004}%
  \BibitemOpen
  \bibfield  {author} {\bibinfo {author} {\bibfnamefont {K.}~\bibnamefont
  {Togano}}, \bibinfo {author} {\bibfnamefont {P.}~\bibnamefont {Badica}},
  \bibinfo {author} {\bibfnamefont {Y.}~\bibnamefont {Nakamori}}, \bibinfo
  {author} {\bibfnamefont {S.}~\bibnamefont {Orimo}}, \bibinfo {author}
  {\bibfnamefont {H.}~\bibnamefont {Takeya}}, \ and\ \bibinfo {author}
  {\bibfnamefont {K.}~\bibnamefont {Hirata}},\ }\href {\doibase
  10.1103/PhysRevLett.93.247004} {\bibfield  {journal} {\bibinfo  {journal}
  {Phys. Rev. Lett.}\ }\textbf {\bibinfo {volume} {93}},\ \bibinfo {pages}
  {247004} (\bibinfo {year} {2004})}\BibitemShut {NoStop}%
\bibitem [{\citenamefont {Badica}\ \emph {et~al.}(2005)\citenamefont {Badica},
  \citenamefont {Kondo},\ and\ \citenamefont {Togano}}]{Badica:2005}%
  \BibitemOpen
  \bibfield  {author} {\bibinfo {author} {\bibfnamefont {P.}~\bibnamefont
  {Badica}}, \bibinfo {author} {\bibfnamefont {T.}~\bibnamefont {Kondo}}, \
  and\ \bibinfo {author} {\bibfnamefont {K.}~\bibnamefont {Togano}},\ }\href
  {\doibase 10.1143/JPSJ.74.1014} {\bibfield  {journal} {\bibinfo  {journal}
  {Journal of the Physical Society of Japan}\ }\textbf {\bibinfo {volume}
  {74}},\ \bibinfo {pages} {1014} (\bibinfo {year} {2005})}\BibitemShut
  {NoStop}%
\bibitem [{\citenamefont {Yuan}\ \emph {et~al.}(2006)\citenamefont {Yuan},
  \citenamefont {Agterberg}, \citenamefont {Hayashi}, \citenamefont {Badica},
  \citenamefont {Vandervelde}, \citenamefont {Togano}, \citenamefont
  {Sigrist},\ and\ \citenamefont {Salamon}}]{Yuan:2006}%
  \BibitemOpen
  \bibfield  {author} {\bibinfo {author} {\bibfnamefont {H.~Q.}\ \bibnamefont
  {Yuan}}, \bibinfo {author} {\bibfnamefont {D.~F.}\ \bibnamefont {Agterberg}},
  \bibinfo {author} {\bibfnamefont {N.}~\bibnamefont {Hayashi}}, \bibinfo
  {author} {\bibfnamefont {P.}~\bibnamefont {Badica}}, \bibinfo {author}
  {\bibfnamefont {D.}~\bibnamefont {Vandervelde}}, \bibinfo {author}
  {\bibfnamefont {K.}~\bibnamefont {Togano}}, \bibinfo {author} {\bibfnamefont
  {M.}~\bibnamefont {Sigrist}}, \ and\ \bibinfo {author} {\bibfnamefont
  {M.~B.}\ \bibnamefont {Salamon}},\ }\href {\doibase
  10.1103/PhysRevLett.97.017006} {\bibfield  {journal} {\bibinfo  {journal}
  {Phys. Rev. Lett.}\ }\textbf {\bibinfo {volume} {97}},\ \bibinfo {pages}
  {017006} (\bibinfo {year} {2006})}\BibitemShut {NoStop}%
\bibitem [{\citenamefont {Nishiyama}\ \emph {et~al.}(2007)\citenamefont
  {Nishiyama}, \citenamefont {Inada},\ and\ \citenamefont
  {Zheng}}]{Nishiyama:2007}%
  \BibitemOpen
  \bibfield  {author} {\bibinfo {author} {\bibfnamefont {M.}~\bibnamefont
  {Nishiyama}}, \bibinfo {author} {\bibfnamefont {Y.}~\bibnamefont {Inada}}, \
  and\ \bibinfo {author} {\bibfnamefont {G.-q.}\ \bibnamefont {Zheng}},\ }\href
  {\doibase 10.1103/PhysRevLett.98.047002} {\bibfield  {journal} {\bibinfo
  {journal} {Phys. Rev. Lett.}\ }\textbf {\bibinfo {volume} {98}},\ \bibinfo
  {pages} {047002} (\bibinfo {year} {2007})}\BibitemShut {NoStop}%
\bibitem [{\citenamefont {Yogi}\ \emph {et~al.}(2004)\citenamefont {Yogi},
  \citenamefont {Kitaoka}, \citenamefont {Hashimoto}, \citenamefont {Yasuda},
  \citenamefont {Settai}, \citenamefont {Matsuda}, \citenamefont {Haga},
  \citenamefont {{\ifmmode \bar{O}\else \={O}\fi{}nuki}}, \citenamefont
  {Rogl},\ and\ \citenamefont {Bauer}}]{Yogi:2004}%
  \BibitemOpen
  \bibfield  {author} {\bibinfo {author} {\bibfnamefont {M.}~\bibnamefont
  {Yogi}}, \bibinfo {author} {\bibfnamefont {Y.}~\bibnamefont {Kitaoka}},
  \bibinfo {author} {\bibfnamefont {S.}~\bibnamefont {Hashimoto}}, \bibinfo
  {author} {\bibfnamefont {T.}~\bibnamefont {Yasuda}}, \bibinfo {author}
  {\bibfnamefont {R.}~\bibnamefont {Settai}}, \bibinfo {author} {\bibfnamefont
  {T.~D.}\ \bibnamefont {Matsuda}}, \bibinfo {author} {\bibfnamefont
  {Y.}~\bibnamefont {Haga}}, \bibinfo {author} {\bibfnamefont {Y.}~\bibnamefont
  {{\ifmmode \bar{O}\else \={O}\fi{}nuki}}}, \bibinfo {author} {\bibfnamefont
  {P.}~\bibnamefont {Rogl}}, \ and\ \bibinfo {author} {\bibfnamefont
  {E.}~\bibnamefont {Bauer}},\ }\href {\doibase 10.1103/PhysRevLett.93.027003}
  {\bibfield  {journal} {\bibinfo  {journal} {Phys. Rev. Lett.}\ }\textbf
  {\bibinfo {volume} {93}},\ \bibinfo {pages} {027003} (\bibinfo {year}
  {2004})}\BibitemShut {NoStop}%
\bibitem [{\citenamefont {Izawa}\ \emph {et~al.}(2005)\citenamefont {Izawa},
  \citenamefont {Kasahara}, \citenamefont {Matsuda}, \citenamefont {Behnia},
  \citenamefont {Yasuda}, \citenamefont {Settai},\ and\ \citenamefont
  {Onuki}}]{Izawa:2005}%
  \BibitemOpen
  \bibfield  {author} {\bibinfo {author} {\bibfnamefont {K.}~\bibnamefont
  {Izawa}}, \bibinfo {author} {\bibfnamefont {Y.}~\bibnamefont {Kasahara}},
  \bibinfo {author} {\bibfnamefont {Y.}~\bibnamefont {Matsuda}}, \bibinfo
  {author} {\bibfnamefont {K.}~\bibnamefont {Behnia}}, \bibinfo {author}
  {\bibfnamefont {T.}~\bibnamefont {Yasuda}}, \bibinfo {author} {\bibfnamefont
  {R.}~\bibnamefont {Settai}}, \ and\ \bibinfo {author} {\bibfnamefont
  {Y.}~\bibnamefont {Onuki}},\ }\href {\doibase 10.1103/PhysRevLett.94.197002}
  {\bibfield  {journal} {\bibinfo  {journal} {Phys. Rev. Lett.}\ }\textbf
  {\bibinfo {volume} {94}},\ \bibinfo {pages} {197002} (\bibinfo {year}
  {2005})}\BibitemShut {NoStop}%
\bibitem [{\citenamefont {Bonalde}\ \emph {et~al.}(2009)\citenamefont
  {Bonalde}, \citenamefont {Ribeiro}, \citenamefont {Br{\"a}mer-Escamilla},
  \citenamefont {Rojas}, \citenamefont {Bauer}, \citenamefont {Prokofiev},
  \citenamefont {Haga}, \citenamefont {Yasuda},\ and\ \citenamefont
  {{\=O}nuki}}]{Bonalde:2009}%
  \BibitemOpen
  \bibfield  {author} {\bibinfo {author} {\bibfnamefont {I.}~\bibnamefont
  {Bonalde}}, \bibinfo {author} {\bibfnamefont {R.~L.}\ \bibnamefont
  {Ribeiro}}, \bibinfo {author} {\bibfnamefont {W.}~\bibnamefont
  {Br{\"a}mer-Escamilla}}, \bibinfo {author} {\bibfnamefont {C.}~\bibnamefont
  {Rojas}}, \bibinfo {author} {\bibfnamefont {E.}~\bibnamefont {Bauer}},
  \bibinfo {author} {\bibfnamefont {A.}~\bibnamefont {Prokofiev}}, \bibinfo
  {author} {\bibfnamefont {Y.}~\bibnamefont {Haga}}, \bibinfo {author}
  {\bibfnamefont {T.}~\bibnamefont {Yasuda}}, \ and\ \bibinfo {author}
  {\bibfnamefont {Y.}~\bibnamefont {{\=O}nuki}},\ }\href
  {http://stacks.iop.org/1367-2630/11/i=5/a=055054} {\bibfield  {journal}
  {\bibinfo  {journal} {New Journal of Physics}\ }\textbf {\bibinfo {volume}
  {11}},\ \bibinfo {pages} {055054} (\bibinfo {year} {2009})}\BibitemShut
  {NoStop}%
\bibitem [{\citenamefont {Kimura}\ \emph {et~al.}(2005)\citenamefont {Kimura},
  \citenamefont {Ito}, \citenamefont {Saitoh}, \citenamefont {Umeda},
  \citenamefont {Aoki},\ and\ \citenamefont {Terashima}}]{Kimura:2005}%
  \BibitemOpen
  \bibfield  {author} {\bibinfo {author} {\bibfnamefont {N.}~\bibnamefont
  {Kimura}}, \bibinfo {author} {\bibfnamefont {K.}~\bibnamefont {Ito}},
  \bibinfo {author} {\bibfnamefont {K.}~\bibnamefont {Saitoh}}, \bibinfo
  {author} {\bibfnamefont {Y.}~\bibnamefont {Umeda}}, \bibinfo {author}
  {\bibfnamefont {H.}~\bibnamefont {Aoki}}, \ and\ \bibinfo {author}
  {\bibfnamefont {T.}~\bibnamefont {Terashima}},\ }\href {\doibase
  10.1103/PhysRevLett.95.247004} {\bibfield  {journal} {\bibinfo  {journal}
  {Phys. Rev. Lett.}\ }\textbf {\bibinfo {volume} {95}},\ \bibinfo {pages}
  {247004} (\bibinfo {year} {2005})}\BibitemShut {NoStop}%
\bibitem [{\citenamefont {Sugitani}\ \emph {et~al.}(2006)\citenamefont
  {Sugitani}, \citenamefont {Okuda}, \citenamefont {Shishido}, \citenamefont
  {Yamada}, \citenamefont {Thamizhavel}, \citenamefont {Yamamoto},
  \citenamefont {Matsuda}, \citenamefont {Haga}, \citenamefont {Takeuchi},
  \citenamefont {Settai},\ and\ \citenamefont {{\=O}nuki}}]{Sugitani:2006}%
  \BibitemOpen
  \bibfield  {author} {\bibinfo {author} {\bibfnamefont {I.}~\bibnamefont
  {Sugitani}}, \bibinfo {author} {\bibfnamefont {Y.}~\bibnamefont {Okuda}},
  \bibinfo {author} {\bibfnamefont {H.}~\bibnamefont {Shishido}}, \bibinfo
  {author} {\bibfnamefont {T.}~\bibnamefont {Yamada}}, \bibinfo {author}
  {\bibfnamefont {A.}~\bibnamefont {Thamizhavel}}, \bibinfo {author}
  {\bibfnamefont {E.}~\bibnamefont {Yamamoto}}, \bibinfo {author}
  {\bibfnamefont {T.~D.}\ \bibnamefont {Matsuda}}, \bibinfo {author}
  {\bibfnamefont {Y.}~\bibnamefont {Haga}}, \bibinfo {author} {\bibfnamefont
  {T.}~\bibnamefont {Takeuchi}}, \bibinfo {author} {\bibfnamefont
  {R.}~\bibnamefont {Settai}}, \ and\ \bibinfo {author} {\bibfnamefont
  {Y.}~\bibnamefont {{\=O}nuki}},\ }\href {\doibase 10.1143/JPSJ.75.043703}
  {\bibfield  {journal} {\bibinfo  {journal} {Journal of the Physical Society
  of Japan}\ }\textbf {\bibinfo {volume} {75}},\ \bibinfo {pages} {043703}
  (\bibinfo {year} {2006})}\BibitemShut {NoStop}%
\bibitem [{\citenamefont {Mukuda}\ \emph {et~al.}(2008)\citenamefont {Mukuda},
  \citenamefont {Fujii}, \citenamefont {Ohara}, \citenamefont {Harada},
  \citenamefont {Yashima}, \citenamefont {Kitaoka}, \citenamefont {Okuda},
  \citenamefont {Settai},\ and\ \citenamefont {Onuki}}]{Mukuda:2008}%
  \BibitemOpen
  \bibfield  {author} {\bibinfo {author} {\bibfnamefont {H.}~\bibnamefont
  {Mukuda}}, \bibinfo {author} {\bibfnamefont {T.}~\bibnamefont {Fujii}},
  \bibinfo {author} {\bibfnamefont {T.}~\bibnamefont {Ohara}}, \bibinfo
  {author} {\bibfnamefont {A.}~\bibnamefont {Harada}}, \bibinfo {author}
  {\bibfnamefont {M.}~\bibnamefont {Yashima}}, \bibinfo {author} {\bibfnamefont
  {Y.}~\bibnamefont {Kitaoka}}, \bibinfo {author} {\bibfnamefont
  {Y.}~\bibnamefont {Okuda}}, \bibinfo {author} {\bibfnamefont
  {R.}~\bibnamefont {Settai}}, \ and\ \bibinfo {author} {\bibfnamefont
  {Y.}~\bibnamefont {Onuki}},\ }\href {\doibase 10.1103/PhysRevLett.100.107003}
  {\bibfield  {journal} {\bibinfo  {journal} {Phys. Rev. Lett.}\ }\textbf
  {\bibinfo {volume} {100}},\ \bibinfo {pages} {107003} (\bibinfo {year}
  {2008})}\BibitemShut {NoStop}%
\bibitem [{\citenamefont {Bauer}\ \emph {et~al.}(2010)\citenamefont {Bauer},
  \citenamefont {Rogl}, \citenamefont {Chen}, \citenamefont {Khan},
  \citenamefont {Michor}, \citenamefont {Hilscher}, \citenamefont {Royanian},
  \citenamefont {Kumagai}, \citenamefont {Li}, \citenamefont {Li},
  \citenamefont {Podloucky},\ and\ \citenamefont {Rogl}}]{Bauer:2010}%
  \BibitemOpen
  \bibfield  {author} {\bibinfo {author} {\bibfnamefont {E.}~\bibnamefont
  {Bauer}}, \bibinfo {author} {\bibfnamefont {G.}~\bibnamefont {Rogl}},
  \bibinfo {author} {\bibfnamefont {X.-Q.}\ \bibnamefont {Chen}}, \bibinfo
  {author} {\bibfnamefont {R.~T.}\ \bibnamefont {Khan}}, \bibinfo {author}
  {\bibfnamefont {H.}~\bibnamefont {Michor}}, \bibinfo {author} {\bibfnamefont
  {G.}~\bibnamefont {Hilscher}}, \bibinfo {author} {\bibfnamefont
  {E.}~\bibnamefont {Royanian}}, \bibinfo {author} {\bibfnamefont
  {K.}~\bibnamefont {Kumagai}}, \bibinfo {author} {\bibfnamefont {D.~Z.}\
  \bibnamefont {Li}}, \bibinfo {author} {\bibfnamefont {Y.~Y.}\ \bibnamefont
  {Li}}, \bibinfo {author} {\bibfnamefont {R.}~\bibnamefont {Podloucky}}, \
  and\ \bibinfo {author} {\bibfnamefont {P.}~\bibnamefont {Rogl}},\ }\href
  {\doibase 10.1103/PhysRevB.82.064511} {\bibfield  {journal} {\bibinfo
  {journal} {Phys. Rev. B}\ }\textbf {\bibinfo {volume} {82}},\ \bibinfo
  {pages} {064511} (\bibinfo {year} {2010})}\BibitemShut {NoStop}%
\bibitem [{\citenamefont {Karki}\ \emph {et~al.}(2010)\citenamefont {Karki},
  \citenamefont {Xiong}, \citenamefont {Vekhter}, \citenamefont {Browne},
  \citenamefont {Adams}, \citenamefont {Young}, \citenamefont {Thomas},
  \citenamefont {Chan}, \citenamefont {Kim},\ and\ \citenamefont
  {Prozorov}}]{Karki:2010}%
  \BibitemOpen
  \bibfield  {author} {\bibinfo {author} {\bibfnamefont {A.~B.}\ \bibnamefont
  {Karki}}, \bibinfo {author} {\bibfnamefont {Y.~M.}\ \bibnamefont {Xiong}},
  \bibinfo {author} {\bibfnamefont {I.}~\bibnamefont {Vekhter}}, \bibinfo
  {author} {\bibfnamefont {D.}~\bibnamefont {Browne}}, \bibinfo {author}
  {\bibfnamefont {P.~W.}\ \bibnamefont {Adams}}, \bibinfo {author}
  {\bibfnamefont {D.~P.}\ \bibnamefont {Young}}, \bibinfo {author}
  {\bibfnamefont {K.~R.}\ \bibnamefont {Thomas}}, \bibinfo {author}
  {\bibfnamefont {J.~Y.}\ \bibnamefont {Chan}}, \bibinfo {author}
  {\bibfnamefont {H.}~\bibnamefont {Kim}}, \ and\ \bibinfo {author}
  {\bibfnamefont {R.}~\bibnamefont {Prozorov}},\ }\href {\doibase
  10.1103/PhysRevB.82.064512} {\bibfield  {journal} {\bibinfo  {journal} {Phys.
  Rev. B}\ }\textbf {\bibinfo {volume} {82}},\ \bibinfo {pages} {064512}
  (\bibinfo {year} {2010})}\BibitemShut {NoStop}%
\bibitem [{\citenamefont {Dresselhaus}(1955)}]{Dresselhaus:1995}%
  \BibitemOpen
  \bibfield  {author} {\bibinfo {author} {\bibfnamefont {G.}~\bibnamefont
  {Dresselhaus}},\ }\href {\doibase 10.1103/PhysRev.100.580} {\bibfield
  {journal} {\bibinfo  {journal} {Phys. Rev.}\ }\textbf {\bibinfo {volume}
  {100}},\ \bibinfo {pages} {580} (\bibinfo {year} {1955})}\BibitemShut
  {NoStop}%
\bibitem [{\citenamefont {Rashba}(1960)}]{Rashba:1960}%
  \BibitemOpen
  \bibfield  {author} {\bibinfo {author} {\bibfnamefont {E.~I.}\ \bibnamefont
  {Rashba}},\ }\href@noop {} {\bibfield  {journal} {\bibinfo  {journal} {Sov.
  Phys. Solid State}\ }\textbf {\bibinfo {volume} {2}},\ \bibinfo {pages}
  {1109} (\bibinfo {year} {1960})}\BibitemShut {NoStop}%
\bibitem [{\citenamefont {Hu}\ \emph {et~al.}(2011)\citenamefont {Hu},
  \citenamefont {Jiang}, \citenamefont {Liu},\ and\ \citenamefont
  {Pu}}]{Hui:2011}%
  \BibitemOpen
  \bibfield  {author} {\bibinfo {author} {\bibfnamefont {H.}~\bibnamefont
  {Hu}}, \bibinfo {author} {\bibfnamefont {L.}~\bibnamefont {Jiang}}, \bibinfo
  {author} {\bibfnamefont {X.-J.}\ \bibnamefont {Liu}}, \ and\ \bibinfo
  {author} {\bibfnamefont {H.}~\bibnamefont {Pu}},\ }\href {\doibase
  10.1103/PhysRevLett.107.195304} {\bibfield  {journal} {\bibinfo  {journal}
  {Phys. Rev. Lett.}\ }\textbf {\bibinfo {volume} {107}},\ \bibinfo {pages}
  {195304} (\bibinfo {year} {2011})}\BibitemShut {NoStop}%
\bibitem [{\citenamefont {Yu}\ and\ \citenamefont {Zhai}(2011)}]{Yu:2011}%
  \BibitemOpen
  \bibfield  {author} {\bibinfo {author} {\bibfnamefont {Z.-Q.}\ \bibnamefont
  {Yu}}\ and\ \bibinfo {author} {\bibfnamefont {H.}~\bibnamefont {Zhai}},\
  }\href {\doibase 10.1103/PhysRevLett.107.195305} {\bibfield  {journal}
  {\bibinfo  {journal} {Phys. Rev. Lett.}\ }\textbf {\bibinfo {volume} {107}},\
  \bibinfo {pages} {195305} (\bibinfo {year} {2011})}\BibitemShut {NoStop}%
\bibitem [{\citenamefont {Dell'Anna}\ \emph {et~al.}(2012)\citenamefont
  {Dell'Anna}, \citenamefont {Mazzarella},\ and\ \citenamefont
  {Salasnich}}]{DellAnna:2012}%
  \BibitemOpen
  \bibfield  {author} {\bibinfo {author} {\bibfnamefont {L.}~\bibnamefont
  {Dell'Anna}}, \bibinfo {author} {\bibfnamefont {G.}~\bibnamefont
  {Mazzarella}}, \ and\ \bibinfo {author} {\bibfnamefont {L.}~\bibnamefont
  {Salasnich}},\ }\href {\doibase 10.1103/PhysRevA.86.053632} {\bibfield
  {journal} {\bibinfo  {journal} {Phys. Rev. A}\ }\textbf {\bibinfo {volume}
  {86}},\ \bibinfo {pages} {053632} (\bibinfo {year} {2012})}\BibitemShut
  {NoStop}%
\bibitem [{\citenamefont {Li}\ \emph {et~al.}(2012)\citenamefont {Li},
  \citenamefont {Covaci},\ and\ \citenamefont {Marsiglio}}]{Li:2012}%
  \BibitemOpen
  \bibfield  {author} {\bibinfo {author} {\bibfnamefont {Z.}~\bibnamefont
  {Li}}, \bibinfo {author} {\bibfnamefont {L.}~\bibnamefont {Covaci}}, \ and\
  \bibinfo {author} {\bibfnamefont {F.}~\bibnamefont {Marsiglio}},\ }\href
  {\doibase 10.1103/PhysRevB.85.205112} {\bibfield  {journal} {\bibinfo
  {journal} {Phys. Rev. B}\ }\textbf {\bibinfo {volume} {85}},\ \bibinfo
  {pages} {205112} (\bibinfo {year} {2012})}\BibitemShut {NoStop}%
\bibitem [{\citenamefont {He}\ and\ \citenamefont {Huang}(2012)}]{He:2012}%
  \BibitemOpen
  \bibfield  {author} {\bibinfo {author} {\bibfnamefont {L.}~\bibnamefont
  {He}}\ and\ \bibinfo {author} {\bibfnamefont {X.-G.}\ \bibnamefont {Huang}},\
  }\href {\doibase 10.1103/PhysRevLett.108.145302} {\bibfield  {journal}
  {\bibinfo  {journal} {Phys. Rev. Lett.}\ }\textbf {\bibinfo {volume} {108}},\
  \bibinfo {pages} {145302} (\bibinfo {year} {2012})}\BibitemShut {NoStop}%
\bibitem [{\citenamefont {Ganichev}\ \emph {et~al.}(2004)\citenamefont
  {Ganichev}, \citenamefont {Bel'kov}, \citenamefont {Golub}, \citenamefont
  {Ivchenko}, \citenamefont {Schneider}, \citenamefont {Giglberger},
  \citenamefont {Eroms}, \citenamefont {{De Boeck}}, \citenamefont {Borghs},
  \citenamefont {Wegscheider}, \citenamefont {Weiss},\ and\ \citenamefont
  {Prettl}}]{Ganichev:2004}%
  \BibitemOpen
  \bibfield  {author} {\bibinfo {author} {\bibfnamefont {S.~D.}\ \bibnamefont
  {Ganichev}}, \bibinfo {author} {\bibfnamefont {V.~V.}\ \bibnamefont
  {Bel'kov}}, \bibinfo {author} {\bibfnamefont {L.~E.}\ \bibnamefont {Golub}},
  \bibinfo {author} {\bibfnamefont {E.~L.}\ \bibnamefont {Ivchenko}}, \bibinfo
  {author} {\bibfnamefont {P.}~\bibnamefont {Schneider}}, \bibinfo {author}
  {\bibfnamefont {S.}~\bibnamefont {Giglberger}}, \bibinfo {author}
  {\bibfnamefont {J.}~\bibnamefont {Eroms}}, \bibinfo {author} {\bibfnamefont
  {J.}~\bibnamefont {{De Boeck}}}, \bibinfo {author} {\bibfnamefont
  {G.}~\bibnamefont {Borghs}}, \bibinfo {author} {\bibfnamefont
  {W.}~\bibnamefont {Wegscheider}}, \bibinfo {author} {\bibfnamefont
  {D.}~\bibnamefont {Weiss}}, \ and\ \bibinfo {author} {\bibfnamefont
  {W.}~\bibnamefont {Prettl}},\ }\href {\doibase 10.1103/PhysRevLett.92.256601}
  {\bibfield  {journal} {\bibinfo  {journal} {Phys. Rev. Lett.}\ }\textbf
  {\bibinfo {volume} {92}},\ \bibinfo {pages} {256601} (\bibinfo {year}
  {2004})}\BibitemShut {NoStop}%
\bibitem [{\citenamefont {Koralek}\ \emph {et~al.}(2009)\citenamefont
  {Koralek}, \citenamefont {Weber}, \citenamefont {Orenstein}, \citenamefont
  {Bernevig}, \citenamefont {Zhang}, \citenamefont {Mack},\ and\ \citenamefont
  {Awschalom}}]{Koralek:2009}%
  \BibitemOpen
  \bibfield  {author} {\bibinfo {author} {\bibfnamefont {J.~D.}\ \bibnamefont
  {Koralek}}, \bibinfo {author} {\bibfnamefont {C.~P.}\ \bibnamefont {Weber}},
  \bibinfo {author} {\bibfnamefont {J.}~\bibnamefont {Orenstein}}, \bibinfo
  {author} {\bibfnamefont {B.~A.}\ \bibnamefont {Bernevig}}, \bibinfo {author}
  {\bibfnamefont {S.-c.}\ \bibnamefont {Zhang}}, \bibinfo {author}
  {\bibfnamefont {S.}~\bibnamefont {Mack}}, \ and\ \bibinfo {author}
  {\bibfnamefont {D.~D.}\ \bibnamefont {Awschalom}},\ }\href {\doibase
  10.1038/nature07871} {\bibfield  {journal} {\bibinfo  {journal} {Nature}\
  }\textbf {\bibinfo {volume} {458}},\ \bibinfo {pages} {610} (\bibinfo {year}
  {2009})}\BibitemShut {NoStop}%
\bibitem [{\citenamefont {Yan}\ and\ \citenamefont {Gu}(2014)}]{Yan:2014}%
  \BibitemOpen
  \bibfield  {author} {\bibinfo {author} {\bibfnamefont {X.}~\bibnamefont
  {Yan}}\ and\ \bibinfo {author} {\bibfnamefont {Q.}~\bibnamefont {Gu}},\
  }\href {\doibase 10.1016/j.ssc.2014.02.013} {\bibfield  {journal} {\bibinfo
  {journal} {Solid State Communications}\ }\textbf {\bibinfo {volume} {187}},\
  \bibinfo {pages} {68} (\bibinfo {year} {2014})}\BibitemShut {NoStop}%
\bibitem [{\citenamefont {Dias}\ \emph {et~al.}(2016)\citenamefont {Dias},
  \citenamefont {Frota},\ and\ \citenamefont {Ghosh}}]{Dias:2016}%
  \BibitemOpen
  \bibfield  {author} {\bibinfo {author} {\bibfnamefont {C.~O.}\ \bibnamefont
  {Dias}}, \bibinfo {author} {\bibfnamefont {H.~O.}\ \bibnamefont {Frota}}, \
  and\ \bibinfo {author} {\bibfnamefont {A.}~\bibnamefont {Ghosh}},\ }\href
  {\doibase 10.1002/pssb.201552557} {\bibfield  {journal} {\bibinfo  {journal}
  {physica status solidi (b)}\ }\textbf {\bibinfo {volume} {253}},\ \bibinfo
  {pages} {1824} (\bibinfo {year} {2016})}\BibitemShut {NoStop}%
\bibitem [{\citenamefont {Volpez}\ \emph {et~al.}(2017)\citenamefont {Volpez},
  \citenamefont {Loss},\ and\ \citenamefont {Klinovaja}}]{Volpez:2017}%
  \BibitemOpen
  \bibfield  {author} {\bibinfo {author} {\bibfnamefont {Y.}~\bibnamefont
  {Volpez}}, \bibinfo {author} {\bibfnamefont {D.}~\bibnamefont {Loss}}, \ and\
  \bibinfo {author} {\bibfnamefont {J.}~\bibnamefont {Klinovaja}},\ }\href
  {\doibase 10.1103/PhysRevB.96.085422} {\bibfield  {journal} {\bibinfo
  {journal} {Phys. Rev. B}\ }\textbf {\bibinfo {volume} {96}},\ \bibinfo
  {pages} {085422} (\bibinfo {year} {2017})}\BibitemShut {NoStop}%
\bibitem [{\citenamefont {Xiao}\ \emph {et~al.}(2012)\citenamefont {Xiao},
  \citenamefont {Liu}, \citenamefont {Feng}, \citenamefont {Xu},\ and\
  \citenamefont {Yao}}]{DiXiao:2012}%
  \BibitemOpen
  \bibfield  {author} {\bibinfo {author} {\bibfnamefont {D.}~\bibnamefont
  {Xiao}}, \bibinfo {author} {\bibfnamefont {G.-B.}\ \bibnamefont {Liu}},
  \bibinfo {author} {\bibfnamefont {W.}~\bibnamefont {Feng}}, \bibinfo {author}
  {\bibfnamefont {X.}~\bibnamefont {Xu}}, \ and\ \bibinfo {author}
  {\bibfnamefont {W.}~\bibnamefont {Yao}},\ }\href {\doibase
  10.1103/PhysRevLett.108.196802} {\bibfield  {journal} {\bibinfo  {journal}
  {Phys. Rev. Lett.}\ }\textbf {\bibinfo {volume} {108}},\ \bibinfo {pages}
  {196802} (\bibinfo {year} {2012})}\BibitemShut {NoStop}%
\bibitem [{\citenamefont {Tafti}\ \emph {et~al.}(2013)\citenamefont {Tafti},
  \citenamefont {Fujii}, \citenamefont {Juneau-Fecteau}, \citenamefont
  {{Ren{\'e} de Cotret}}, \citenamefont {Doiron-Leyraud}, \citenamefont
  {Asamitsu},\ and\ \citenamefont {Taillefer}}]{Tafti:2013}%
  \BibitemOpen
  \bibfield  {author} {\bibinfo {author} {\bibfnamefont {F.~F.}\ \bibnamefont
  {Tafti}}, \bibinfo {author} {\bibfnamefont {T.}~\bibnamefont {Fujii}},
  \bibinfo {author} {\bibfnamefont {A.}~\bibnamefont {Juneau-Fecteau}},
  \bibinfo {author} {\bibfnamefont {S.}~\bibnamefont {{Ren{\'e} de Cotret}}},
  \bibinfo {author} {\bibfnamefont {N.}~\bibnamefont {Doiron-Leyraud}},
  \bibinfo {author} {\bibfnamefont {A.}~\bibnamefont {Asamitsu}}, \ and\
  \bibinfo {author} {\bibfnamefont {L.}~\bibnamefont {Taillefer}},\ }\href
  {\doibase 10.1103/PhysRevB.87.184504} {\bibfield  {journal} {\bibinfo
  {journal} {Phys. Rev. B}\ }\textbf {\bibinfo {volume} {87}},\ \bibinfo
  {pages} {184504} (\bibinfo {year} {2013})}\BibitemShut {NoStop}%
\bibitem [{\citenamefont {Sau}\ \emph {et~al.}(2011)\citenamefont {Sau},
  \citenamefont {Sensarma}, \citenamefont {Powell}, \citenamefont {Spielman},\
  and\ \citenamefont {{Das Sarma}}}]{Sau:2011}%
  \BibitemOpen
  \bibfield  {author} {\bibinfo {author} {\bibfnamefont {J.~D.}\ \bibnamefont
  {Sau}}, \bibinfo {author} {\bibfnamefont {R.}~\bibnamefont {Sensarma}},
  \bibinfo {author} {\bibfnamefont {S.}~\bibnamefont {Powell}}, \bibinfo
  {author} {\bibfnamefont {I.~B.}\ \bibnamefont {Spielman}}, \ and\ \bibinfo
  {author} {\bibfnamefont {S.}~\bibnamefont {{Das Sarma}}},\ }\href {\doibase
  10.1103/PhysRevB.83.140510} {\bibfield  {journal} {\bibinfo  {journal} {Phys.
  Rev. B}\ }\textbf {\bibinfo {volume} {83}},\ \bibinfo {pages} {140510}
  (\bibinfo {year} {2011})}\BibitemShut {NoStop}%
\bibitem [{\citenamefont {Jiang}\ \emph {et~al.}(2011)\citenamefont {Jiang},
  \citenamefont {Liu}, \citenamefont {Hu},\ and\ \citenamefont
  {Pu}}]{Jiang:2011}%
  \BibitemOpen
  \bibfield  {author} {\bibinfo {author} {\bibfnamefont {L.}~\bibnamefont
  {Jiang}}, \bibinfo {author} {\bibfnamefont {X.-J.}\ \bibnamefont {Liu}},
  \bibinfo {author} {\bibfnamefont {H.}~\bibnamefont {Hu}}, \ and\ \bibinfo
  {author} {\bibfnamefont {H.}~\bibnamefont {Pu}},\ }\href {\doibase
  10.1103/PhysRevA.84.063618} {\bibfield  {journal} {\bibinfo  {journal} {Phys.
  Rev. A}\ }\textbf {\bibinfo {volume} {84}},\ \bibinfo {pages} {063618}
  (\bibinfo {year} {2011})}\BibitemShut {NoStop}%
\bibitem [{\citenamefont {{Hsu Yi-Ting}}\ \emph {et~al.}(2017)\citenamefont
  {{Hsu Yi-Ting}}, \citenamefont {{Vaezi Abolhassan}}, \citenamefont {{Fischer
  Mark H.}},\ and\ \citenamefont {{Kim Eun-Ah}}}]{Hsu:2017}%
  \BibitemOpen
  \bibfield  {author} {\bibinfo {author} {\bibnamefont {{Hsu Yi-Ting}}},
  \bibinfo {author} {\bibnamefont {{Vaezi Abolhassan}}}, \bibinfo {author}
  {\bibnamefont {{Fischer Mark H.}}}, \ and\ \bibinfo {author} {\bibnamefont
  {{Kim Eun-Ah}}},\ }\href {\doibase 10.1038/ncomms14985;;;;;;
  10.1038/ncomms14985} {\bibfield  {journal} {\bibinfo  {journal} {Nature
  Communications}\ }\textbf {\bibinfo {volume} {8}},\ \bibinfo {pages} {14985}
  (\bibinfo {year} {2017})}\BibitemShut {NoStop}%
\bibitem [{\citenamefont {{Xi Xiaoxiang}}\ \emph {et~al.}(2015)\citenamefont
  {{Xi Xiaoxiang}}, \citenamefont {{Wang Zefang}}, \citenamefont {{Zhao
  Weiwei}}, \citenamefont {{Park Ju-Hyun}}, \citenamefont {{Law Kam Tuen}},
  \citenamefont {{Berger Helmuth}}, \citenamefont {{Forr{\'o} L{\'a}szl{\'o}}},
  \citenamefont {{Shan Jie}},\ and\ \citenamefont {{Mak Kin Fai}}}]{Xi:2015}%
  \BibitemOpen
  \bibfield  {author} {\bibinfo {author} {\bibnamefont {{Xi Xiaoxiang}}},
  \bibinfo {author} {\bibnamefont {{Wang Zefang}}}, \bibinfo {author}
  {\bibnamefont {{Zhao Weiwei}}}, \bibinfo {author} {\bibnamefont {{Park
  Ju-Hyun}}}, \bibinfo {author} {\bibnamefont {{Law Kam Tuen}}}, \bibinfo
  {author} {\bibnamefont {{Berger Helmuth}}}, \bibinfo {author} {\bibnamefont
  {{Forr{\'o} L{\'a}szl{\'o}}}}, \bibinfo {author} {\bibnamefont {{Shan Jie}}},
  \ and\ \bibinfo {author} {\bibnamefont {{Mak Kin Fai}}},\ }\href {\doibase
  10.1038/nphys3538;;;;; 10.1038/nphys3538} {\bibfield  {journal} {\bibinfo
  {journal} {Nature Physics}\ }\textbf {\bibinfo {volume} {12}},\ \bibinfo
  {pages} {139} (\bibinfo {year} {2015})}\BibitemShut {NoStop}%
\bibitem [{\citenamefont {Schliemann}(2006)}]{SCHLIEMANN:2006aa}%
  \BibitemOpen
  \bibfield  {author} {\bibinfo {author} {\bibfnamefont {J.}~\bibnamefont
  {Schliemann}},\ }\href {\doibase 10.1142/S021797920603370X} {\bibfield
  {journal} {\bibinfo  {journal} {International Journal of Modern Physics B}\
  }\textbf {\bibinfo {volume} {20}},\ \bibinfo {pages} {1015} (\bibinfo {year}
  {2006})}\BibitemShut {NoStop}%
\bibitem [{\citenamefont {Winkler}(2003)}]{Winkler:2003}%
  \BibitemOpen
  \bibfield  {author} {\bibinfo {author} {\bibfnamefont {R.}~\bibnamefont
  {Winkler}},\ }\enquote {\bibinfo {title} {{Spin-Orbit Coupling Effects in
  Two-Dimensional Electron and Hole Systems}},}\ \ (\bibinfo  {publisher}
  {Springer Tracts in Modern Physics},\ \bibinfo {year} {2003})\ Chap.\
  \bibinfo {chapter} {Innovation and Intellectual Property Rights}\BibitemShut
  {NoStop}%
\bibitem [{\citenamefont {{Ganichev}}\ and\ \citenamefont
  {{Golub}}(2014)}]{Ganichev:2014}%
  \BibitemOpen
  \bibfield  {author} {\bibinfo {author} {\bibfnamefont {S.~D.}\ \bibnamefont
  {{Ganichev}}}\ and\ \bibinfo {author} {\bibfnamefont {L.~E.}\ \bibnamefont
  {{Golub}}},\ }\href {\doibase 10.1002/pssb.201350261} {\bibfield  {journal}
  {\bibinfo  {journal} {Physica Status Solidi B Basic Research}\ }\textbf
  {\bibinfo {volume} {251}},\ \bibinfo {pages} {1801} (\bibinfo {year}
  {2014})}\BibitemShut {NoStop}%
\bibitem [{\citenamefont {Vas{\rq}ko}(1979)}]{Vasco:1979}%
  \BibitemOpen
  \bibfield  {author} {\bibinfo {author} {\bibfnamefont {F.~T.}\ \bibnamefont
  {Vas{\rq}ko}},\ }\href@noop {} {\bibfield  {journal} {\bibinfo  {journal}
  {Pis{\rq}ma Zh. Eksp. Teor. Fiz}\ }\textbf {\bibinfo {volume} {30}},\
  \bibinfo {pages} {574} (\bibinfo {year} {1979})}\BibitemShut {NoStop}%
\bibitem [{\citenamefont {Bychkov}\ and\ \citenamefont
  {Rashba}(1984)}]{Bychkov:1984}%
  \BibitemOpen
  \bibfield  {author} {\bibinfo {author} {\bibfnamefont {Y.~A.}\ \bibnamefont
  {Bychkov}}\ and\ \bibinfo {author} {\bibfnamefont {E.~I.}\ \bibnamefont
  {Rashba}},\ }\href@noop {} {\bibfield  {journal} {\bibinfo  {journal}
  {Pis{\rq}ma Zh. Eksp. Teor. Fiz}\ }\textbf {\bibinfo {volume} {39}},\
  \bibinfo {pages} {66} (\bibinfo {year} {1984})}\BibitemShut {NoStop}%
\bibitem [{\citenamefont {Engels}\ \emph {et~al.}(1997)\citenamefont {Engels},
  \citenamefont {Lange}, \citenamefont {Sch{\"a}pers},\ and\ \citenamefont
  {L{\"u}th}}]{Engels:1997}%
  \BibitemOpen
  \bibfield  {author} {\bibinfo {author} {\bibfnamefont {G.}~\bibnamefont
  {Engels}}, \bibinfo {author} {\bibfnamefont {J.}~\bibnamefont {Lange}},
  \bibinfo {author} {\bibfnamefont {T.}~\bibnamefont {Sch{\"a}pers}}, \ and\
  \bibinfo {author} {\bibfnamefont {H.}~\bibnamefont {L{\"u}th}},\ }\href
  {\doibase 10.1103/PhysRevB.55.R1958} {\bibfield  {journal} {\bibinfo
  {journal} {Phys. Rev. B}\ }\textbf {\bibinfo {volume} {55}},\ \bibinfo
  {pages} {R1958} (\bibinfo {year} {1997})}\BibitemShut {NoStop}%
\bibitem [{\citenamefont {Nitta}\ \emph {et~al.}(1997)\citenamefont {Nitta},
  \citenamefont {Akazaki}, \citenamefont {Takayanagi},\ and\ \citenamefont
  {Enoki}}]{Nitta:1997}%
  \BibitemOpen
  \bibfield  {author} {\bibinfo {author} {\bibfnamefont {J.}~\bibnamefont
  {Nitta}}, \bibinfo {author} {\bibfnamefont {T.}~\bibnamefont {Akazaki}},
  \bibinfo {author} {\bibfnamefont {H.}~\bibnamefont {Takayanagi}}, \ and\
  \bibinfo {author} {\bibfnamefont {T.}~\bibnamefont {Enoki}},\ }\href
  {\doibase 10.1103/PhysRevLett.78.1335} {\bibfield  {journal} {\bibinfo
  {journal} {Phys. Rev. Lett.}\ }\textbf {\bibinfo {volume} {78}},\ \bibinfo
  {pages} {1335} (\bibinfo {year} {1997})}\BibitemShut {NoStop}%
\bibitem [{\citenamefont {Frigeri}\ \emph {et~al.}(2004)\citenamefont
  {Frigeri}, \citenamefont {Agterberg}, \citenamefont {Koga},\ and\
  \citenamefont {Sigrist}}]{Frigeri:2004}%
  \BibitemOpen
  \bibfield  {author} {\bibinfo {author} {\bibfnamefont {P.~A.}\ \bibnamefont
  {Frigeri}}, \bibinfo {author} {\bibfnamefont {D.~F.}\ \bibnamefont
  {Agterberg}}, \bibinfo {author} {\bibfnamefont {A.}~\bibnamefont {Koga}}, \
  and\ \bibinfo {author} {\bibfnamefont {M.}~\bibnamefont {Sigrist}},\ }\href
  {\doibase 10.1103/PhysRevLett.92.097001} {\bibfield  {journal} {\bibinfo
  {journal} {Phys. Rev. Lett.}\ }\textbf {\bibinfo {volume} {92}},\ \bibinfo
  {pages} {097001} (\bibinfo {year} {2004})}\BibitemShut {NoStop}%
\bibitem [{\citenamefont {Dresselhaus}\ \emph {et~al.}(2008)\citenamefont
  {Dresselhaus}, \citenamefont {Dresselhaus},\ and\ \citenamefont
  {Jorio}}]{Dresselhaus:2008aa}%
  \BibitemOpen
  \bibfield  {author} {\bibinfo {author} {\bibfnamefont {M.~S.}\ \bibnamefont
  {Dresselhaus}}, \bibinfo {author} {\bibfnamefont {G.}~\bibnamefont
  {Dresselhaus}}, \ and\ \bibinfo {author} {\bibfnamefont {A.}~\bibnamefont
  {Jorio}},\ }\href@noop {} {\emph {\bibinfo {title} {{Group theory :
  application to the physics of condensed matter}}}}\ (\bibinfo  {publisher}
  {Springer},\ \bibinfo {address} {Berlin},\ \bibinfo {year}
  {2008})\BibitemShut {NoStop}%
\bibitem [{\citenamefont {Tanaka}\ \emph {et~al.}(2010)\citenamefont {Tanaka},
  \citenamefont {Mizuno}, \citenamefont {Yokoyama}, \citenamefont {Yada},\ and\
  \citenamefont {Sato}}]{Tanaka:2010aa}%
  \BibitemOpen
  \bibfield  {author} {\bibinfo {author} {\bibfnamefont {Y.}~\bibnamefont
  {Tanaka}}, \bibinfo {author} {\bibfnamefont {Y.}~\bibnamefont {Mizuno}},
  \bibinfo {author} {\bibfnamefont {T.}~\bibnamefont {Yokoyama}}, \bibinfo
  {author} {\bibfnamefont {K.}~\bibnamefont {Yada}}, \ and\ \bibinfo {author}
  {\bibfnamefont {M.}~\bibnamefont {Sato}},\ }\href {\doibase
  10.1103/PhysRevLett.105.097002} {\bibfield  {journal} {\bibinfo  {journal}
  {Phys. Rev. Lett.}\ }\textbf {\bibinfo {volume} {105}},\ \bibinfo {pages}
  {097002} (\bibinfo {year} {2010})}\BibitemShut {NoStop}%
\bibitem [{\citenamefont {Schnyder}\ \emph {et~al.}(2008)\citenamefont
  {Schnyder}, \citenamefont {Ryu}, \citenamefont {Furusaki},\ and\
  \citenamefont {Ludwig}}]{Schnyder:2008}%
  \BibitemOpen
  \bibfield  {author} {\bibinfo {author} {\bibfnamefont {A.~P.}\ \bibnamefont
  {Schnyder}}, \bibinfo {author} {\bibfnamefont {S.}~\bibnamefont {Ryu}},
  \bibinfo {author} {\bibfnamefont {A.}~\bibnamefont {Furusaki}}, \ and\
  \bibinfo {author} {\bibfnamefont {A.~W.~W.}\ \bibnamefont {Ludwig}},\ }\href
  {\doibase 10.1103/PhysRevB.78.195125} {\bibfield  {journal} {\bibinfo
  {journal} {Phys. Rev. B}\ }\textbf {\bibinfo {volume} {78}},\ \bibinfo
  {pages} {195125} (\bibinfo {year} {2008})}\BibitemShut {NoStop}%
\bibitem [{\citenamefont {Chiu}\ \emph {et~al.}(2016)\citenamefont {Chiu},
  \citenamefont {Teo}, \citenamefont {Schnyder},\ and\ \citenamefont
  {Ryu}}]{Schnyder:2016}%
  \BibitemOpen
  \bibfield  {author} {\bibinfo {author} {\bibfnamefont {C.-K.}\ \bibnamefont
  {Chiu}}, \bibinfo {author} {\bibfnamefont {J.~C.~Y.}\ \bibnamefont {Teo}},
  \bibinfo {author} {\bibfnamefont {A.~P.}\ \bibnamefont {Schnyder}}, \ and\
  \bibinfo {author} {\bibfnamefont {S.}~\bibnamefont {Ryu}},\ }\href {\doibase
  10.1103/RevModPhys.88.035005} {\bibfield  {journal} {\bibinfo  {journal}
  {Rev. Mod. Phys.}\ }\textbf {\bibinfo {volume} {88}},\ \bibinfo {pages}
  {035005} (\bibinfo {year} {2016})}\BibitemShut {NoStop}%
\end{thebibliography}%
\end{document}